\documentstyle[aps,preprint,epsf]{revtex}
\newcommand{\beq}{\begin{equation}}
\newcommand{\eeq}{\end{equation}}
\begin{document}
\draft
\tightenlines
\title{ Appearance of Fermion Condensate in Fermi-Liquids}
\author{V.R. Shaginyan
\footnote{E--mail: vrshag@thd.pnpi.spb.ru}}
\address{Petersburg Nuclear Physics Institute,
Gatchina 188300, Russia}
\maketitle\begin{abstract}
The appearance of the fermion condensation, which can be compared
to the Bose-Einstein condensation, in high-$T_c$ metals, heavy
fermion metals and in other liquids is considered, its properties are
discussed and  a large number of experimental evidences in favor of
the existence of the fermion condensate (FC) is presented. The
appearance of FC is a signature of  the fermion condensation quantum
phase transition (FCQPT), that separates the regions of normal and
strongly correlated electron liquids. Beyond the FCQPT point  the
quasiparticle system is divided into two subsystems, one containing
normal quasiparticles and the other - FC localized at the Fermi level.
In the superconducting state the quasiparticle dispersion in systems
with FC can be represented by two straight lines, characterized by
effective masses $M^*_{FC}$ and $M^*_L$, and intersecting near the
binding energy $E_0$, which is of the order of the superconducting
gap. The same quasiparticles and the energy scale $E_0$
persist in the normal state. Fermion systems
with FC have features of a ``quantum protectorate".
Electron systems with FC, which exhibit large
deviations from the Landau Fermi liquid behavior, can be driven into
the Landau Fermi liquid by applying a magnetic field $B$.
Thus, the essence of strongly correlated electron liquids can be
controlled by magnetic fields.
A re-entrance into the strongly correlated
regime is observed if the magnetic field $(B-B_{c0})$ decreases to zero,
while the effective mass $M^*$ diverges as
$M^*\propto1/\sqrt{B-B_{c0}}$.
The regime is restored at some temperature
$T^*\propto\sqrt{B-B_{c0}}$. In some cases the critical
field $B_{c0}\to 0$. The behavior of Fermi systems approaching
FCQPT from the disordered phase is considered.
This behavior can be viewed as a highly correlated one.
The essence of highly correlated electron liquids can be
also controlled by magnetic fields.
A re-entrance into the highly correlated
regime is observed if $(B-B_{c0})$ decreases to zero,
while $M^*$ diverges as $M^*\propto(B-B_c)^{-2/3}$.
The regime is restored at some temperature
$T^*\propto(B-B_{c0})^{4/3}$.
The observed $B-T$ phase diagrams have a strong impact on the
magnetoresistance of strongly and highly correlated electron liquids.
We expect that FCQPT takes place in trapped Fermi gases and in
a low density neutron matter. When the system recedes from FCQPT
it becomes conventional Landau Fermi liquid.
\end{abstract}


\section{Introduction}

In a system of interacting bosons at temperatures lower than the
temperature of Bose-Einstein condensation \cite{bel1,bel2},
a finite number of
particles is concentrated in the lowest level. In case of a
noninteracting Bose gas at the zero temperature, $T=0$, this
number is simply equal to the total number of particles in the
system. In a homogeneous system of noninteracting Bosons, the
lowest level is the state with zero momentum, and the ground state
energy is equal to zero. For a noninteracting Fermi system such a
state is impossible, and its ground state energy $E_{gs}$
reduces to the kinetic energy and is proportional to the
total number of particles. Imagine an interacting system of
fermions with a pure repulsive interaction. Let us increase its
interaction strength. As soon as it becomes sufficiently large and the
potential energy starts to prevail over the kinetic energy, we can
expect the system to undergo a phase transition when a finite
number of the Cooper like pairs with an infinitely small binding
energy can condensate at the Fermi level. Such a state resembles
the Bose-Einstein condensation and can be viewed as fermion
condensation \cite{yaf}. This phase transition leads to the onset of the
fermion condensate (FC) and separates a strongly interacting Fermi
liquid from a strongly correlated one. Lowering the potential
energy, the fermion condensation decreases the total energy.
Unlike the Bose-Einstein condensation,
which occurs  even in a system of noninteracting bosons, the
fermion condensation can take place if the coupling constant of
the interaction is large, or the corresponding Landau amplitudes are
large and repulsive.

One of the most challenging problems of modern physics is the
structure and properties of systems with large coupling constants.
It is well-known that a theory of liquids with strong interaction
is close to the problem of systems with a big coupling constant.
The first solution to this problem was offered by the Landau
theory of Fermi liquids, later called "normal", by introducing the
notion of quasiparticles and parameters, which characterize the
effective interaction among them \cite{lan}.
The Landau theory can be viewed as the low energy
effective theory in which high energy
degrees of freedom are removed
at the cost of introducing the effective
interaction parameters. Usually, it is
assumed that the stability of the ground state of a Landau liquid is
determined by the Pomeranchuk stability conditions:
the stability is violated when even one of the Landau effective
interaction parameters
is negative and reaches a critical value. Note
that the new phase, new ground state, at which the stability
conditions are restored can in principle be again described within the
framework of the same theory.

It has been demonstrated, however rather recently \cite{ks} that
the Pomeranchuk conditions cover not all possible instabilities:
one of them is missed. It corresponds to the situation when, at
the temperature $T=0$, the effective mass, the most important
characteristic of Landau quasiparticles, can become infinitely
large. Such a situation, leading to profound consequences, can
take place when the corresponding Landau amplitude
being repulsive reaches some critical value.
This leads to a completely new class of strongly correlated Fermi
liquids with FC \cite{ks,vol}, which is separated
from that of a normal
Fermi liquid by the fermion condensation quantum phase transition
(FCQPT) \cite{ms}.

In the FCQPT case we are dealing with the strong coupling limit
where an absolutely reliable answer cannot be given on the bases
of pure theoretical first principle foundation. Therefore, the
only way to verify that FC occurs is to consider experimental
facts, which can be interpreted as confirming the existence of
such a state. We believe that these facts are seen in some
features of those two-dimensional (2D) systems with interacting
electrons or holes, which can be represented by doped quantum wells
and high-$T_c$ superconductors. Considering the heavy-fermion
metals, the 2D systems of $^3$He, the trapped neutrons and Fermi
gases, we will show that FC can exist also in these systems.

The goal of our paper is to describe the behavior of Fermi systems
with FC and to show that the existing data on strongly correlated
liquids can be well understood within the theory of Fermi liquids
with FC. In Sec. II, we review the general features of Fermi
liquids with FC in their normal state. Sec. III is devoted to
consideration of the superconductivity in the presence of FC. We
show that the superconducting state is totally transformed by the
presence of FC. For instance, the maximum value $\Delta_1$ of the
superconducting gap can be as large as $\Delta_1\sim 0.1
\varepsilon_F$, while for normal superconductors one has
$\Delta_1\sim 10^{-3} \varepsilon_F$.
Here $\varepsilon_F$ is the Fermi level.
In Sec. IV we describe the
quasiparticle's dispersion and its lineshape and show that they
strongly deviate from the case of normal Landau liquids. In Sections V
and VII we apply our theory to explain the main properties of
heavy-fermion metals and the field induced Landau Fermi liquid
behavior observed in these metals.
The obtained $B-T$ phase diagrams have a strong impact on the
magnetoresistance of both strongly and highly correlated electron
liquids. We demonstrate that it is possible
to control the main properties, or even the essence,  of both strongly
and highly correlated electron liquids by weak magnetic fields. Sec.
VI deals with the possibility of FCQPT in different Fermi systems, such
as 2D systems of electrons and 2D $^3$He liquids, neutron matter at low
density and trapped Fermi gases.  In Sec. VII we describe the behavior
of Fermi systems which approach FCQPT from the disordered phase and
show that this behavior is observed in heavy-fermion metals. In the
vicinity of FCQPT, this behavior can be viewed as a highly correlated
one, because the effective mass is large and strongly depends on the
density, temperature and magnetic fields. Finally, in Sec. VIII, we
summarize our main results.

\section {Normal state of Fermi liquids with FC}

Let us start by explaining the important points of the FC theory,
which is a special solution of the Fermi-liquid theory equations
\cite{lan} for the quasiparticle occupation numbers $n(p,T)$,
\beq
\frac{\delta(F-\mu N)}{\delta n(p,T)}\ =\ \varepsilon(p,T)
-\mu(T)-T\ln\frac{1-n(p,T)}{n(p,T)}\ =\ 0\ , \eeq which depends on
the momentum $p$ and temperature $T$. Here $F=E-TS$ is the free
energy, $S$ is the entropy, and $\mu$ is the chemical potential,
while \beq \varepsilon(p,T)\ =\ \frac{\delta E[n(p,T)]}{\delta
n(p,T)}\ , \eeq is the quasiparticle energy. This energy is a
functional of $n(p,T)$ just like the total energy $E[n(p,T)]$,
entropy $S[n(p,T)]$, and other thermodynamic functions. The
entropy $S[n(p,T)]$ is given by the familiar expression
$$
S[n(p,T)]=-2\sum_{p}\left[n(p,T)\ln n(p,T)
+(1-n(p,T))\ln(1-n(p,T))\right],
$$
which stems from purely combinatorial considerations. Eq. (1) is
usually presented as the Fermi-Dirac distribution \beq n(p,T)\ =\
\left\{1+\exp\left[\frac{(\varepsilon(p,T)-\mu)}
{T}\right]\right\}^{-1}. \eeq At $T\to 0$, one
gets from Eqs. (1) and (3) the standard
solution $n_F(p,T\to0)\to\theta(p_F-p)$, with
$\varepsilon(p\simeq p_F)-\mu=p_F(p-p_F)/M^*_L$, where $p_F$ is
the Fermi momentum and $M^*_L$ is the Landau effective mass
\cite{lan} \beq \frac1{M^*_L}\ =\
\frac1p\,\frac{d\varepsilon(p,T=0)}{dp}|_{p=p_F}\ .\eeq It is
implied that $M^*_L$ is positive and finite at the Fermi momentum
$p_F$. As a result, the $T$-dependent corrections to $M^*_L$, to
the quasiparticle energy $\varepsilon (p)$, and to other
quantities, start with $T^2$-terms. But this solution is not the
only one possible. There exist also "anomalous" solutions of Eq. (1)
associated with the so-called fermion condensation \cite{ks,ksk}.
Being continuous and satisfying the inequality $0<n(p)<1$ within
some region in $p$, such solutions $n(p)$ admit a finite value for
the logarithm in Eq. (1) at $T\rightarrow 0$ yielding \beq
\varepsilon(p)\ =\ \frac{\delta E[n(p)]} {\delta n(p)}\ =\ \mu\ ;
\quad p_i\ \le\ p \leq p_f. \eeq At $T=0$, Eq. (5) determines
FCQPT, possessing solutions at some density $x=x_{FC}$ as
soon as the effective inter-fermion interaction becomes
sufficiently strong \cite{ksk,ksz}. For instance, in the ordinary
electron liquid, the effective inter-electron interaction is
proportional to the dimensionless average interparticle distance
$r_s=r_0/a_B$, with $r_0\sim 1/p_F$ being the average distance
and $a_B$ is the Bohr radius.
When fermion condensation can take place at $1\ll r_{s}$,
it is considered to be in the low density electron liquid \cite{ksz}.
At $x\leq x_{FC}$ the new state of electron liquid with FC
is defined by Eq. (5) and characterized by a flat part of
the spectrum in the $(p_i-p_f)$ region. Apparently, the momenta $p_i$
and $p_f$ have to satisfy $p_i<p_F<p_f$.
Note, that a formation of the flat part of the
spectrum has been recently confirmed in Ref. \cite{dzyal,irkh}.

Equation (5) leads to the minimal value of $E$, as a functional of
$n(p)$, when a strong rearrangement of the single particle
spectra can take place in the system under consideration.
We see from Eq. (5)
that the occupation numbers $n(p)$ become variational parameters:
the FC solution appears if the energy $E$ can be lowered by
alteration of the occupation numbers $n(p)$. Thus, within the region
$p_i<p<p_f$, the solution $n(p)=n_F(p)+\delta n(p)$ deviates from
the Fermi step function $n_F(p)$ in such a way that the energy
$\varepsilon(p)$ stays constant while outside this region $n(p)$
coincides with $n_F(p)$. It is essential to note that the
general consideration presented above has been verified by
inspecting some simple models. As a result, it was shown that the
onset of the FC does lead to lowering of the free energy
\cite{ksk,dkss}.

It follows from the above consideration that the superconductivity
order parameter $\kappa({\bf p})=\sqrt{n({\bf p})(1-n({\bf p}))}$
has a nonzero value over the region occupied by FC. The
superconducting gap $\Delta({\bf p})$ being linear in the coupling
constant of the particle-particle interaction $V({\bf p}_1,{\bf
p}_2)$ increases the value of $T_c$ because one has $2T_c\simeq
\Delta_1$ \cite{dkss} within the standard Bardeen-Cooper-Schrieffer
(BCS) theory \cite{bcs}. As shown in Sec. III, if the
superconducting gap is non-zero, $\Delta_1\neq 0$, the FC
quasiparticle effective mass $M^*_{FC}$ becomes finite.  Consequently,
the density of states of these quasiparticles at the Fermi level
becomes finite and the quasiparticles involved are delocalized. On the
other hand, even at $T=0$, $\Delta_1$ can vanish, provided the
interparticle interaction $V({\bf p}_1,{\bf p}_2)$ is either repulsive
or absent.  Then, as seen from Eq. (5), the Landau quasiparticle system
becomes separated into two subsystems. The first contains the Landau
quasiparticles, while the second, related to FC, is localized at the
Fermi surface and is formed by dispersionless quasiparticles. As a
result, beyond the point of the FC phase transition the standard
Kohn-Sham scheme for the single-particle equations is no longer valid
\cite{vsl}. Such a behavior of systems with FC is clearly different
from what one expects from the well known local density approach.
Therefore, this in generally a very powerful method is hardly
applicable for the description of systems with FC. It is also seen from
Eq. (5) that a system with FC has a well-defined Fermi surface.

Let us assume that with the decrease of the density or growth of the
interaction strength FC has just taken place. It means that
$p_i\to p_f\to p_F$, and the deviation $\delta n(p)$ is small.
Expanding the functional $E[n(p)]$ in Taylor's series with respect
to $\delta n(p)$ and retaining the leading terms, one obtains from
Eq. (5) the following relation \beq \mu\ =\ \varepsilon({\bf p})\
=\ \varepsilon_0({\bf p})+\int F_L({\bf p},{\bf p}_1)\delta n({\bf
p_1}) \frac{d{\bf p}_1}{(2\pi)^2}\ ; \quad p_i\leq p \leq p_f\ ,
\eeq where $F_L({\bf p},{\bf p}_1)=\delta^2 E/\delta n({\bf
p})\delta n({\bf p}_1)$ is the Landau effective interaction. Both
quantities, the interaction and the single-particle energy
$\varepsilon_0(p)$ are calculated at $n(p)=n_F(p)$. Equation (6)
acquires nontrivial solutions at some density
$x=x_{FC}$ and FCQPT takes
place if the Landau amplitudes depending on the density are
positive and sufficiently large, so that the potential energy is
bigger than the kinetic energy. Then the transformation of the
Fermi step function $n(p)=\theta(p_F-p)$ into the smooth function
defined by Eq. (5) becomes possible \cite{ks,ksk}. It is seen from
Eq. (5) that the FC quasiparticles form a collective state, since
their energies are defined by the macroscopical number of
quasiparticles within the momentum region $p_i-p_f$. The shape of
the excitation spectra related to FC is not affected by the Landau
interaction, which, generally speaking, depends on the system's
properties, including the collective states, impurities, etc. The
only thing determined by the interaction is the width of the FC
region $p_i-p_f$ provided the interaction is sufficiently strong
to produce the FC phase transition at all. Thus, we can conclude
that the spectra related to FC are of a universal form, being
dependent, as we will see below, mainly on temperature $T$ if
$T>T_c$  or on the superconducting gap at $T<T_c$.

According to Eq. (1), the single-particle energy
$\varepsilon(p,T)$
within the interval $p_i-p_f$ at $T_c\leq T\ll
T_f$ is linear in $T$ \cite{kcs}. At the Fermi level,
one obtains by expanding $\ln(...)$ in
terms of $n(p)$ \beq \varepsilon(p,T)-\mu(T)\ =\
T\ln\frac{1-n(p)}{n(p)}\ \simeq\ T\frac{1-2n(p)}{n(p)}\bigg|_{p\simeq
p_F}\ . \eeq Here $T_f$ is the temperature, above which FC effects
become insignificant \cite{dkss}, \beq \frac{T_f}{\varepsilon_F}\ \sim\
\frac{p_f^2-p_i^2}{2M\varepsilon_F}\ \sim\
\frac{\Omega_{FC}}{\Omega_F}\ . \eeq In this formula $\Omega_{FC}$
is the FC volume, $\varepsilon_F$ is the Fermi energy, and
$\Omega_F$ is the volume of the Fermi sphere. We note that at
$T_c\leq T\ll T_f$ the occupation numbers $n(p)$ are approximately
independent of $T$, being given by Eq. (5). At finite
temperatures according to Eq.  (1), the dispersionless plateau
$\varepsilon(p)=\mu$ is slightly turned counter-clockwise about $\mu$.
As a result, the plateau is just a little tilted and rounded off at the
end points. According to Eq. (7) the effective mass $M^*_{FC}$ related
to FC is given by, \beq M^*_{FC}\ \simeq\ p_F\frac{p_f-p_i}{4T}\ . \eeq
To obtain Eq. (9) an approximation for the derivative $dn(p)/dp\simeq
-1/(p_f-p_i)$ was used.

Having in mind that $(p_f-p_i)\ll p_F$ and using Eqs. (8) and (9),
the following estimates for the effective mass $M^*_{FC}$ are
obtained: \beq \frac{M^*_{FC}}{M}\ \sim\ \frac{N(0)}{N_0(0)}\
\sim\ \frac{T_f}T\ . \eeq Eqs. (9) and (10) show the temperature
dependence of $M^*_{FC}$. In Eq. (10) $M$ denotes the bare
electron mass, $N_0(0)$ is the density of states of noninteracting
electron gas, and $N(0)$ is the density of states at the Fermi
level. Multiplying both sides of Eq. (9) by $(p_f-p_i)$ we obtain
the energy scale $E_0$ separating the slow dispersing low energy
part related to the effective mass $M^*_{FC}$ from the faster
dispersing relatively high energy part defined by the effective
mass $M^*_{L}$ \cite{ms,ars}, \beq E_0\ \simeq\ 4T\ . \eeq It is
seen from Eq. (11) that the scale $E_0$ does not depend on the
condensate volume. The single particle excitations are defined
according to Eq. (9) by the temperature and by $(p_f-p_i)$,
given by Eq. (5). Thus, we conclude that the one-electron spectrum
is negligible disturbed by thermal excitations, impurities, etc,
which are the features of the "quantum protectorate"
\cite{rlp,pa}.

It is pertinent to note that outside the FC region the single
particle spectrum is not affected by the temperature, being
defined by $M^*_L$. Thus we come to the conclusion that a system
with FC is characterized by two effective masses: $M^*_{FC}$ which
is related to the single particle spectrum at lower energy scale
and $M^*_L$ describing the spectrum at higher energy scale.  The
existence of two effective masses is manifested by a break (or kink)
in the quasiparticle dispersion, which can be approximated by two
straight lines intersecting at the energy $E_0$.  This break takes
place at temperatures $T_c\leq T\ll T_f$, in accord with the
experimental data \cite{blk}, and, as we will see, at $T\leq T_c$
which is also in accord with the experimental facts \cite{blk,krc}.
The quasiparticle formalism is applicable to this problem since
the width $\gamma$ of single particle excitations is not large
compared to their energy, being proportional to the temperature,
$\gamma\sim T$ at $T>T_c$ \cite{dkss}. The lineshape can be
approximated by a simple Lorentzian \cite{ars}, consistent with
experimental data obtained from scans at a constant binding energy
\cite{vall} (see Sec. IV).

It is seen from Eq. (5) that at the point of FC phase transition
$p_f\to p_i\to p_F$, $M^*_{FC}$ and the density of states, as it
follows from Eqs. (5) and (10), tend to infinity. One can conclude
that at $T=0$ and as soon as $x\to x_{FC}$, FCQPT takes place
being connected to the absolute growth of $M^*_{L}$.

It is essential to have in mind, that the onset of the charge
density wave instability in a many-electron system, such as
an electron liquid, which takes place as soon as the effective
inter-electron constant reaches its critical value $r_s=r_{cdw}$,
is preceded by the unlimited growth of the effective mass, see Sec. VI.
Therefore the FC occurs before the onset of the charge density
wave. Hence, at $T=0$, when $r_s$ reaches its critical value
$r_{FC}$ corresponding to $x_{FC}$,
$r_{FC}<r_{cdw}$, FCQPT inevitably takes place \cite{ksz}. It is
pertinent to note that this growth of the effective mass with
decreasing electron density was observed experimentally in a
metallic 2D electron system in silicon at $r_s\simeq 7.5$
\cite{skdk}. Therefore we can take $r_{FC}\sim 7.5$. On the other
hand, there exist charge density waves or strong fluctuations of
charge ordering in underdoped high-$T_c$ superconductors
\cite{grun}. Thus the formation of FC in high-$T_c$ compounds can
be thought as a general property of an electron liquid of low
density which is embedded in these solids rather than an uncommon
and anomalous solution of Eq. (1) \cite{ksz}. Beyond the point of
FCQPT, the condensate volume is proportional to $(r_s-r_{FC})$ as
well as $T_f/\varepsilon_F\sim (r_s-r_{FC})/r_{FC}$ at least when
$(r_s-r_{FC})/r_{FC}\ll 1$, and we obtain \beq
\frac{r_s-r_{FC}}{r_{FC}}\sim \frac{p_f-p_{i}}{p_{F}}
\sim \frac{x_{FC}-x}{x_{FC}}. \eeq FC
serves as a stimulator that creates new phase transitions, which
lift the degeneration of the spectrum. For example FC can generate
spin density waves or antiferromagnetic phase transition, thus
leading to a whole variety of new properties of the system under
consideration. Then, the onset of the charge density wave is
preceded by FCQPT, and both of these phases can coexist at the
sufficiently low density when $r_s\geq r_{cdw}$.

We have demonstrated above that superconductivity is strongly
aided by FC because both of the phases are characterized by the
same order parameter. As a result, the superconductivity removing
the spectrum degeneration, "wins" the competition with the other
phase transitions up to the critical temperature $T_c$. We
now turn to the consideration of the superconducting state and
quasiparticle dispersions  at $T\leq T_c$.

\section{The superconducting state}

At $T=0$, the ground state energy $E_{gs}[\kappa({\bf p}),n({\bf
p})]$ of a 2D electron liquid is a functional of the order parameter
of the superconducting state $\kappa({\bf p})$ and of the
quasiparticle occupation numbers $n({\bf p})$. This energy is
determined by the known equation of the weak-coupling theory of
superconductivity, see e.g. \cite{til} \beq E_{gs}\ =\ E[n({\bf
p})]+\int \lambda_0V({\bf p}_1,{\bf p}_2) \kappa({\bf p}_1)
\kappa^*({\bf p}_2) \frac{d{\bf p}_1d{\bf p}_2}{(2\pi)^4}\ . \eeq
Here  $E[n({\bf p})]$ is the ground-state energy of a normal Fermi
liquid, $n({\bf p})=v^2({\bf p})$ and $\kappa({\bf p})=v({\bf
p})u({\bf p})$. Here $u({\bf p})$ and $v({\bf p})$ are the
normalized coherence factors,  $v^2({\bf p})+u^2({\bf p})=1$.
It is assumed that the pairing
interaction $\lambda_0V({\bf p}_1,{\bf p}_2)$ is weak. Minimizing
$E_{gs}$ with respect to $\kappa({\bf p})$ we obtain the equation
connecting the single-particle energy $\varepsilon({\bf p})$ to
$\Delta({\bf p})$, \beq \varepsilon({\bf p})-\mu\ =\ \Delta({\bf
p}) \frac{1-2v^2({\bf p})} {2\kappa({\bf p})}\ , \eeq where the
single-particle energy $\varepsilon({\bf p})$ is determined by the
Landau equation (2). The equation for the superconducting gap
$\Delta({\bf p})$  takes the form
\begin{eqnarray}
\Delta({\bf p}) &=& -\int\lambda_0V({\bf p},{\bf p}_1)\kappa({\bf p}_1)
\frac{d{\bf p}_1}{4\pi^2} \nonumber\\
&=& -\frac{1}{2}\int\lambda_0 V({\bf p},{\bf p}_1)
\frac{\Delta({\bf p}_1)}{\sqrt{(\varepsilon({\bf p}_1)-\mu)^2
+\Delta^2({\bf p}_1)}} \frac{d{\bf p}_1}{4\pi^2}\ .
\end{eqnarray}
If $\lambda_0\to 0$, then the maximum value $\Delta_1\to 0$ and
Eq. (14) reduces to Eq. (5)  \cite{ks} \beq \varepsilon({\bf
p})-\mu\ =\ 0,\quad \mbox{ if}\quad 0<n({\bf p})<1;\: p_i\leq
p\leq p_f\ . \eeq Now we can study the relationships between the state
defined by Eq. (16), or by Eq. (5), and the superconductivity. At
$T=0$, Eq. (16) defines a particular state of a Fermi-liquid with
FC, for which the modulus of the order parameter $|\kappa({\bf
p})|$ has finite values in the $L_{FC}$ range of momenta $p_i\leq
p\leq p_f$, and $\Delta_1\to 0$ in the $L_{FC}$.
Such a state can
be considered as superconducting, with an infinitely small value
of $\Delta_1$, so that the entropy of this state is equal to zero.
It is obvious that this state being driven by the quantum phase
transition disappears at $T>0$ \cite{ms}. Any quantum phase
transition, which takes place at temperature $T=0$, is determined
by a control parameter other than temperature, for instance, by
pressure, by magnetic field, or by the
density of mobile charge carriers $x\sim
1/r_s^2$. The quantum phase transition occurs at the quantum
critical point. As any phase transition, the quantum phase transition
is related to the order parameter, which induces a broken symmetry. In
our case as we show in Sec. II, the control parameter is the density of
a system, which determines the strength of the Landau effective
interaction, and the order parameter is $\kappa({\bf p})$. As we
point out in Sec. V, the existence of such a state can be revealed
experimentally. Since the order parameter
$\kappa({\bf p})$ is suppressed
by a magnetic field $B$, when $B^2\sim \Delta_1^2$, a weak
magnetic field $B$ will destroy the state with FC converting the
strongly correlated Fermi liquid into the normal Landau Fermi
liquid. In this case the magnetic field play a role of the
control parameter.

When $p_i\to p_F\to p_f$, Eq. (16) determines the critical
density $x_{FC}$ at which FCQPT takes place. It follows from Eq.
(16) that the system becomes divided into two quasiparticle
subsystems: the first subsystem in the $L_{FC}$ range is
characterized  by the quasiparticles with the effective mass
$M^*_{FC}\propto1/\Delta_1$, while the second one is occupied by
quasiparticles with finite mass $M^*_L$ and momenta $p<p_i$. The
density of states near the Fermi level tends to infinity,
$N(0)\sim M^*_{FC}\sim 1/\Delta_1$. The quasiparticles with
$M^*_{FC}$ occupy the same energy level and form pairs with
binding energy of the order of $\Delta_1$ and with average
momentum $p_0$, $p_0/p_F\sim (p_f-p_i)/p_F\ll 1$. Therefore, this
state strongly resembles the Bose-Einstein condensation when
quasiparticles occupy the same energy level. But these have to be
spread over the range $L_{FC}$ in momentum space due to the
exclusion principle. In contrast to the Bose-Einstein
condensation, the fermion condensation temperature  is $T_c=0$.
And in contrast to the ordinary superconductivity, the fermion
condensation is driven by the Landau repulsive interaction rather
than by relatively weak attractive quasiparticle-quasiparticle
interaction $\lambda_0V({\bf p}_1,{\bf p}_2)$.

If $\lambda_0\neq 0$, $\Delta_1$ becomes finite, leading to a
finite value of the effective mass $M^*_{FC}$ in $L_{FC}$, which
can be obtained from Eq. (14) \cite{ms,ars} \beq M^*_{FC}\ \simeq\
p_F\frac{p_f-p_i}{2\Delta_1}\ . \eeq As to the energy scale it is
determined by the parameter $E_0$: \beq E_0\ =\ \varepsilon({\bf
p}_f)-\varepsilon({\bf p}_i)\ \simeq\ 2
\frac{(p_f-p_F)p_F}{M^*_{FC}}\ \simeq\ 2\Delta_1\ . \eeq
It is seen
from Eq. (15) that the superconducting gap depends on the
single-particle spectrum $\varepsilon({\bf p})$. On the other
hand, it follows from Eq. (17) that $\varepsilon({\bf p})$
depends on $\Delta({\bf p})$ provided that at $\Delta_1\to 0$ Eq.
(16) has the solution determining the existence of FC. Let us
assume that $\lambda_0$ is small so that the particle-particle
interaction $\lambda_0 V({\bf p},{\bf p}_1)$ can only lead to a
small perturbation of the order parameter $\kappa({\bf p})$
determined by Eq. (16). It
follows from Eq. (17) that the effective mass and the density of
states $N(0)\propto M^*_{FC}\propto 1/\Delta_1$ are finite.
As a result, we are led to the
conclusion that in contrast to the conventional theory of
superconductivity the single-particle spectrum $\varepsilon({\bf
p})$ strongly depends on the superconducting gap and we have to
solve Eqs. (2), (14) and (15) in a self-consistent way. On the other
hand, let us assume that Eqs. (2) and (15) are solved, and
the effective mass $M^*_{FC}$ is determined. Now one
can fix the dispersion $\varepsilon({\bf p})$ by choosing
the effective mass $M^*$ of system in question equal to
$M^*_{FC}$ and then solve Eq. (15) as it is done in the case of
the conventional theory of superconductivity \cite{bcs}. As a
result, one observes that the superconducting state is
characterized by the Bogoliubov quasiparticles (BQ) \cite{bogol} with
the dispersion
$$E(p)=\sqrt{(\varepsilon(p)-\mu)^2+\Delta^2(p)},$$ and
the normalization condition for the coherence factors $v$ and $u$ is
held.  We are lead to the conclusion that the observed features agree
with the behavior of BQ predicted from BCS theory. This observation
suggests that the superconducting state with FC is BCS-like and
implies the basic validity of BCS formalism in describing the
superconducting state \cite{asjetpl}. It is exactly the case that was
observed experimentally in high-$T_c$ cuprate
Bi$_2$Sr$_2$Ca$_2$Cu$_3$O$_{10+\delta}$ \cite{mat}.

We have returned back to the Landau Fermi liquid theory since high
energy degrees of freedom are eliminated and the quasiparticles
are introduced. The only difference between the Landau Fermi-liquid,
which serves as a basis when constructing the superconducting state,
and Fermi liquid after FCQPT is that we have to expand the number of
relevant low energy degrees of freedom by introducing a new type of
quasiparticles with the effective mass $M^*_{FC}$ given by Eq. (17)
and the energy scale $E_0$ given by Eq. (18). Therefore, the dispersion
$\varepsilon({\bf p})$ is characterized by two effective masses $M^*_L$
and $M^*_{FC}$ and by the scale $E_0$, which define the low temperature
properties including the line shape of quasiparticle excitations
\cite{ms,ams}, while the dispersion of BQ has the standard form.
We note
that both the effective mass $M^*_{FC}$ and the scale $E_0$ are
temperature independent at $T<T_c$, where $T_c$ is the critical
temperature of the superconducting phase transition \cite{ams}.
Obviously, we cannot directly relate these new Landau Fermi-liquid
quasiparticle excitations with the quasiparticle excitations of an
ideal Fermi gas because the system in question has undergone FCQPT.
Nonetheless, the main basis of the Landau Fermi liquid theory survives
FCQPT: the low energy excitations of a strongly correlated liquid with
FC are quasiparticles.

As it was shown above, properties of these new quasiparticles are
closely related to the properties of the superconducting state. We
may say that the quasiparticle system in the range $(p_f-p_i)$
becomes very ``soft'' and is to be considered as a strongly
correlated liquid. On the other hand, the system's properties and
dynamics are dominated by a strong collective effect having its
origin in its proximity to FCQPT and determined by the macroscopic
number of quasiparticles in the range $(p_f-p_i)$. Such a system cannot
be perturbed by the scattering of individual quasiparticles and has
features of a ``quantum protectorate" and demonstrates
the universal behavior \cite{ms,rlp,pa}.  A few remarks related
to the quantum protectorate and the universal behavior \cite{rlp} are
in order here.  As the Landau theory of Fermi liquid, the theory of the
high-temperature superconductivity based on FCQPT deals with the
quasiparticles which are elementary excitations of low energy.  As a
result, this theory produces the general qualitative description of
both the superconducting and normal state.  Of course, one can choose
the phenomenological parameters and obtain the quantitative
consideration of the superconductivity as it can be done in the
framework of the Landau theory when describing a normal Fermi-liquid.
We are led to the conclusion that any theory which is capable of
describing FC and incorporates with the BCS theory will produce the
qualitative picture of the superconducting state and the normal state
which coincides with the picture based on FCQPT.  Both of the pictures
can agree at a numerical level provided the corresponding parameters
are adjusted. For example, since the formation of  flat band is
possible in the Hubbard model \cite{irkh}, generally speaking  one can
repeat the results of theory based on FCQPT in the Hubbard model.  It
is appropriate mention here that the corresponding numerical
description confined to the case of $T=0$ has been obtained within the
Hubbard model \cite{rand}.

Let us assume that the range $L_{FC}$ is small, $(p_f-p_F)/p_F\ll1$,
and $2\Delta_1\ll T_f$ so that the order parameter $\kappa({\bf
p})$ is governed mainly by FC \cite{ms,ams}. To solve Eq. (15)
analytically, we take the Bardeen-Cooper-Schrieffer (BCS)
approximation for the interaction \cite{bcs}:  $\lambda_0V({\bf
p},{\bf p}_1)=-\lambda_0$ if $|\varepsilon({\bf p})-\mu|\leq
\omega_D$, i.e. the interaction is zero outside this region, with
$\omega_D$ being the characteristic phonon energy.  As a result,
the gap becomes dependent only on the temperature, $\Delta({\bf
p})=\Delta_1(T)$, being independent of the momentum, and Eq. (15)
takes the form \beq 1\ =\
N_{FC}\lambda_0\int\limits_0^{E_0/2}\frac{d\xi}
{\sqrt{\xi^2+\Delta^2_1(0)}}
+N_{L}\lambda_0\int\limits_{E_0/2}^{\omega_D}\frac{d\xi}
{\sqrt{\xi^2+\Delta^2_1(0)}}\ . \eeq Here we set
$\xi=\varepsilon({\bf p})-\mu$ and introduce the density of states
$N_{FC}$ in the $L_{FC}$, or $E_0$, range. It follows from Eq.
(17), $N_{FC}=(p_f-p_F)p_F/2\pi\Delta_1(0)$. The density of states
$N_{L}$ in the range $(\omega_D-E_0/2)$ has the standard form
$N_{L}=M^*_{L}/2\pi$. If the energy scale $E_0\to 0$, Eq. (19)
reduces to the BCS equation. On the other hand, assuming that
$E_0\leq2\omega_D$ and omitting the second integral on the right
hand side of Eq. (19), we obtain \beq \Delta_1(0)\ =\
\frac{\lambda_0 p_F(p_f-p_F)}{2\pi}\ln\left(1+\sqrt2\right)\ =\
2\beta\varepsilon_F \frac{p_f-p_F}{p_F}\ln\left(1+\sqrt2\right) ,
\eeq where the Fermi energy $\varepsilon_F=p_F^2/2M^*_L$, and the
dimensionless coupling constant $\beta$ is given by the relation
$\beta=\lambda_0 M^*_L/2\pi$. Taking the usual values of $\beta$ as
$\beta\simeq 0.3$, and assuming
$(p_f-p_F)/p_F\simeq 0.2$, we get
from Eq. (20) a large value of $\Delta_1(0)\sim 0.1\varepsilon_F$,
while for normal metals one has $\Delta_1(0)\sim
10^{-3}\varepsilon_F$. Taking into account the omitted integral,
we obtain \beq \Delta_1(0)\ \simeq\ 2\beta\varepsilon_F
\frac{p_f-p_F}{p_F}\ln\left(1+\sqrt2\right)\left(1+\beta
\ln\frac{2\omega_D}{E_0}\right). \eeq It is seen from Eq. (21)
that the correction due to the second integral is small, provided
$E_0\simeq2\omega_D$. Below we show that $2T_c\simeq \Delta_1(0)$,
which leads to the conclusion that there is no isotope effect
since $\Delta_1$ is independent of $\omega_D$. But this effect is
restored as $E_0\to 0$. Assuming $E_0\sim\omega_D$ and
$E_0>\omega_D$, we see that Eq. (21) has no standard solutions
$\Delta(p)=\Delta_1(T=0)$ because $\omega_D<\varepsilon(p\simeq
p_f)-\mu$ and the interaction vanishes at these momenta. The only
way to obtain solutions is to restore the condition
$E_0<\omega_D$. For instance, we can define such a momentum
$p_D<p_f$ that \beq \Delta_1(0)\ =\ 2\beta\varepsilon_F\
\frac{p_D-p_F}{p_F} \ln\left(1+\sqrt2\right)\ =\ \omega_D\ , \eeq
while the other part in the $L_{FC}$ range can be occupied by a
gap $\Delta_2$ of the different sign, $\Delta_1/\Delta_2<0$. It
follows from Eq. (22) that the isotope effect is preserved, while
both gaps can have $s$-wave symmetry.

At $T\simeq T_c$ Eqs. (17) and (18) are replaced by the equation,
which is valid also at $T_c\leq T\ll T_f$ in accord with Eq. (9)
\cite{ms}: \beq M^*_{FC}\simeq p_F\frac{p_f-p_i}{4T_c},\,\,\,
E_0\simeq 4T_c;\,\,{\mathrm {if}}\,\,T_c\leq T\,
{\mathrm {then}}\,,
M^*_{FC}\simeq p_F\frac{p_f-p_i}{4T},\,\,\, E_0\simeq 4T\ . \eeq
Equation (19) is replaced by its conventional finite temperature
generalization
\begin{eqnarray}
1 & = & N_{FC}\lambda_0\int_0^{E_0/2}
\frac{d\xi}
{\sqrt{\xi^2+\Delta^2_1(T)}}
\tanh\frac{\sqrt{\xi^2+\Delta^2_1(T)}}{2T}\ +
\nonumber\\
& + & N_{L}\lambda_0\int_{E_0/2}^{\omega_D}
\frac{d\xi}{\sqrt{\xi^2+\Delta^2_1(T)}}
\tanh\frac{\sqrt{\xi^2+\Delta^2_1(T)}}{2T}\ .
\end{eqnarray}
Putting $\Delta_1(T\to T_c)\to 0$, we obtain from Eq. (24) \beq
2T_c\ \simeq\ \Delta_1(0)\ , \eeq with $\Delta_1(T=0)$ being given
by Eq. (20). Comparing Eqs. (17), (23) and (25), we see that
$M^*_{FC}$ and $E_0$ are almost temperature independent at $T\leq
T_c$.

Now let us comment about some special features of the superconducting
state with FC. One can define $T_c$ as the temperature when
$\Delta_1(T_c)\equiv 0$. At $T\geq T_c$ Eq. (24) has only the
trivial solution $\Delta_1\equiv 0$. On the other hand, $T_c$ can
be defined as a temperature, at which the superconductivity
disappears. Thus, we have two different definitions, which can
lead to two different temperatures $T_c$ and $T^*$ in case of the
$d$-wave symmetry of the gap. It was shown \cite{ars,sh} that in
the case of the {\bf d}-wave superconductivity in the presence
of FC there is a nontrivial solution of Eq. (24) at $T_c\leq
T\leq T^*$ corresponding to the pseudogap state. It happens when
the gap occupies only such a part of the Fermi surface, which
shrinks as the temperature increases. Here $T^*$ defines the
temperature at which $\Delta_1(T^*)\equiv 0$ and the pseudogap
state vanishes. The superconductivity is destroyed at $T_c$, and
the ratio $2\Delta_1/T_c$ can vary in a wide range and strongly
depends upon the material's properties as it follows from
considerations given in \cite{ars,sh,ms1}. Therefore, if a pseudogap
exists above $T_c$, then $T_c$ is to be replaced by $T^*$ and Eq.
(25) takes the form \beq 2T^*\ \simeq\ \Delta_1(0)\ . \eeq The
ratio $2\Delta_1/T_c$ can reach very high values. For instance, in
the case of Bi$_2$Sr$_2$CaCu$_2$Q$_{6+\delta}$ where the
superconductivity and the pseudogap are considered to be of the
common origin, $2\Delta_1/T_c$ is about 28, while the ratio
$2\Delta_1/T^*\simeq 4$, which is in agreement with the
experimental data for various cuprates \cite{kug}. Note that Eq.
(20) gives also good description of the maximum gap $\Delta_1$ in
the case of the d-wave superconductivity, because the different
regions with the maximum absolute value of $\Delta_1$ and the
maximal density of states can be considered as disconnected
\cite{abr}. Therefore the gap in this region is formed by
attractive phonon interaction, which is approximately independent
of the momenta.

Consider now two possible types of the superconducting gap
$\Delta({\bf p})$ given by Eq. (15) and defined by the interaction
$\lambda_0V({\bf p},{\bf p}_1)$. If this interaction is dominated
by a phonon-mediated attraction, the even solution of Eq. (15)
with the $s$-wave, or the $s+d$ mixed waves will have the lowest
energy. Provided the pairing interaction $\lambda_0V({\bf
p}_1,{\bf p}_2)$ is the combination of both the attractive
interaction and sufficiently strong repulsive interaction, the
$d$-wave odd superconductivity can take place (see e.g.
\cite{abr}). But both the $s$-wave even symmetry and the $d$-wave odd
one lead to approximately the same value of the gap $\Delta_1$ in
Eq.  (21) \cite{ams}. Therefore the non-universal pairing
symmetries in high-$T_c$ superconductivity is likely the result of
the pairing interaction and the $d$-wave pairing symmetry is not
essential. This point of view is supported by the data
\cite{skin,bis,skin1,skin2,chen}. If only the $d$-wave pairing
would exist, the transition from superconducting gap to pseudogap
could take place, so that the superconductivity would be destroyed at
$T_c$, with the superconducting gap being smoothly transformed
into the pseudogap, which closes at some temperature $T^*>T_c$
\cite{sh,ms1}. In the case of the $s$-wave pairing we can expect
the absence of the pseudogap phenomenon in accordance with the
experimental observation (see \cite{chen} and references therein).

We now turn to a consideration of the maximum value of the
superconducting gap $\Delta_1$ as a function of the density $x$ of
the mobile charge carriers. Rewriting in terms of $x\sim r_s^2$
and $x_{FC}\sim r_{FC}^2$ which are
related to the variables $p_i$ and $p_f$
by Eq. (12), Eq. (21) becomes
\beq\Delta_1\propto\beta(x_{FC}-x)x.\eeq Here we take into account
that the Fermi level $\varepsilon_F\propto p_F^2$, the density
$x\propto p_F^2$, and thus, $\varepsilon_F\propto x$. We can
reliably assume that $T_c\propto\Delta_1$ because the empirically
obtained simple bell-shaped curve of $T_c(x)$ in the high temperature
superconductors \cite{vn} should have only a smooth dependence.
Then, $T_c(x)$ in accordance with the data has the form
\cite{ams3} \beq T_c(x)\ \propto\ \beta (x_{FC}-x)x.
\eeq

As an example of the implementation of the previous analysis, let
us consider the main features of a room-temperature
superconductor. The superconductor has to be a quasi
two-dimensional structure like cuprates.
From Eq. (21) it follows, that $\Delta_1\sim \beta \varepsilon_F\propto
\beta/r_s^2$. Noting that FCQPT in 3D systems takes place at
$r_s\sim 20$ and in 2D systems at $r_s\sim 8$ \cite{ksz}, we can
expect that $\Delta_1$ of 3D systems comprises 10\% of the
corresponding maximum value of 2D superconducting gap, reaching a
value as high as 60 meV for underdoped crystals with $T_c=70$
\cite{mzo}. On the other hand, it is seen from Eq. (21), that
$\Delta_1$ can be even large, $\Delta_1\sim 75$ meV, and one can
expect $T_c\sim 300$ K in the case of the $s$ wave pairing as it
follows from the simple relation $2T_c\simeq \Delta_1$.  In fact,
we can safely take $\varepsilon_F\sim 300$ meV, $\beta\sim 0.5$
and $(p_f-p_i)/p_F\sim0.5$. Thus, a possible room-temperature
superconductor has to be the $s$-wave superconductor in order to
get rid of the pseudogap phenomena, which tremendously reduces the
transition temperature. The density $x$ of the mobile charge
carriers must satisfy the condition $x\leq x_{FC}$ and be flexible
to reach the optimal doping level $x_{opt}\simeq x_{FC}/2$.
It is pertinent to note that as it follows from Eqs. (17) and (20) the
effective mass $M^*_{FC}$ does not depend on the momenta $p_F$, $p_f$
and $p_i$.  Then, it is seen from Eqs. (12) and (27) that $M^*_{FC}$
does not depend on $x$. This result is in good agreement with
experimental facts \cite{ino,zhou}. The same is true for the dependence
of the Fermi velocity $v_F=P_F/M^*_{FC}$ on $x$ because the Fermi
momentum $P_F\sim\sqrt{n}$ slightly depends on the electron density
$n=n_0(1-x)$ \cite{ino,zhou}.  Here $n_0$ is the single-particle
electron density at the half-filling.

Now we turn to the calculations of the gap and the specific heat
at the temperatures $T\to T_c$. It is worth noting that this
consideration is valid provided $T^*=T_c$, otherwise the
considered below discontinuity is smoothed out over the
temperature range $T^{*}\div T_c$. For the sake of simplicity, we
calculate the main contribution to the gap and the specific heat
coming from the FC. The function $\Delta_1(T\to T_c)$ is found
from Eq. (24) by expanding the right hand side of the first
integral in powers of $\Delta_1$ and omitting the contribution
from the second integral on the right hand side of Eq. (24). This
procedure leads to the following equation \cite{ams} \beq
\Delta_1(T)\ \simeq\ 3.4T_c\sqrt{1-\frac{T}{T_c}}\ . \eeq Thus,
the gap in the spectrum of the single-particle excitations has
the usual behavior. To calculate the specific heat, the
conventional expression for the entropy $S$ \cite{bcs} can be used
\beq S\ =\ -2\int\left[f({\bf p})\ln f({\bf p}) +(1-f({\bf
p}))\ln(1-f({\bf p}))\right]\frac{d{\bf p}}{(2\pi)^2}\ , \eeq
where \beq f({\bf p})=\frac{1}{1+\exp[E({\bf p})/T]}\ ; \quad
E({\bf p}) =\sqrt{(\varepsilon({\bf p})-\mu)^2+\Delta_1^2(T)}\ .
\eeq The specific heat $C$ is determined by the equation
\begin{eqnarray}
C &=& T\frac{dS}{dT}\ \simeq\ 4\frac{N_{FC}}{T^2}\int\limits_0^{E_0}
f(E)(1-f(E))\left[E^2+T\Delta_1(T)
\frac{d\Delta_1(T)}{dT}\right]d\xi\ +
\nonumber\\
&+& 4\frac{N_{L}}{T^2}\int\limits_{E_0}^{\omega_D}
f(E)(1-f(E))\left[E^2+T\Delta_1(T)\frac{d\Delta_1(T)}{dT}\right]d\xi\ .
\end{eqnarray}
In deriving Eq. (32) we again used the variable $\xi$ and the
densities of states $N_{FC}$ and $N_{L}$, just as before in
connection with Eq. (21), and employed the notation
$E=\sqrt{\xi^2+\Delta_1^2(T)}$. Eq. (32) predicts the conventional
discontinuity $\delta C$ in the specific heat $C$ at $T_c$ because
of the last term in the square brackets of Eq. (32). Using Eq.
(29) to calculate this term and omitting the second integral on
the right hand side of Eq. (32), we obtain \beq \delta\, C\
\simeq\ \frac3{2\pi}\ (p_f-p_i)\,p_F\ . \eeq This is
in contrast to the
conventional result where the discontinuity is a linear function of
$T_c$. $\delta C$ is independent of the critical temperature $T_c$
because as seen from Eq. (23) the density of states varies
inversely with $T_c$. Note, that in deriving Eq. (33) we took into
account the main contribution coming from the FC. This term
vanishes as soon as $E_0\to0$ and the second integral of Eq. (32)
gives the conventional result.

\section{The lineshape of the single-particle spectra}

The lineshape $L(q,\omega)$ of the single-particle
spectrum is a function of two variables. Measurements
carried out at a fixed binding energy $\omega=\omega_0$,
with $\omega_0$ being the energy of a single-particle excitation, determine
the lineshape $L(q,\omega=\omega_0)$ as a function of the momentum $q$.
We have shown above that $M^*_{FC}$ is finite and constant at
$T\leq T_c$. Therefore, at excitation energies $\omega\leq E_0$, the
system behaves like an ordinary superconducting Fermi liquid with the
effective mass given by Eq. (17) \cite{ms,ars}. At $T_c\leq T$ the low
energy effective mass $M^*_{FC}$ is finite and is given by Eq. (9).
Once again, at the energies $\omega<E_0$, the system behaves as a
Fermi liquid, the single-particle spectrum is well defined while the
width of single-particle excitations is of the order of $T$
\cite{ms,dkss}. This behavior was observed in experiments
measuring the lineshape at a fixed energy \cite{vall,feng}.

The lineshape can also be determined as a function
$L(q=q_0,\omega)$ at a fixed $q=q_0$.  At small $\omega$, the
lineshape resembles the one considered above, and $L(q=q_0,\omega)$
has the characteristic maximum and width. At energies $\omega\geq
E_0$, the quasiparticles with the mass $M^*_{L}$ become important,
leading to the increase of $L(q=q_0,\omega)$. As a result, the
function $L(q=q_0,\omega)$ possesses
the known peak-dip-hump structure \cite{dess} directly defined by the
existence of the two effective masses $M^*_{FC}$ and $M^*_L$
\cite{ms,ars}. We can conclude that in contrast to the Landau
quasiparticles, these quasiparticles have a more complicated lineshape.

To develop deeper quantitative and analytical insight into
the problem, we use the Kramers-Kr\"{o}nig transformation to
construct the imaginary part ${\mathrm{Im}}\Sigma({\bf
p},\varepsilon)$ of the self-energy $\Sigma({\bf p},\varepsilon)$
starting with the real one ${\mathrm{Re}}\Sigma({\bf
p},\varepsilon)$, which defines the effective mass \cite{mig} \beq
\left. \frac1{M^*}\ =\ \left(\frac{1}{M}+\frac{1}{p_F}
\frac{\partial{\mathrm{Re}}\Sigma}{\partial p}\right)\right/
\left(1-\frac{\partial{\mathrm{Re}}\Sigma}{\partial
\varepsilon}\right). \eeq Here $M$ is the bare mass, while the
relevant momenta $p$ and energies $\varepsilon$ obey the following
strong inequalities: $|p-p_F|/p_F\ll 1$, and
$\varepsilon/\varepsilon_F\ll 1$. We take
${\mathrm{Re}}\Sigma({\bf p},\varepsilon)$ in the simplest form
which accounts for the change of the effective mass at the energy
scale $E_0$: \beq \mbox{Re }\Sigma({\bf
p},\varepsilon)=-\varepsilon
\frac{M^*_{FC}}M\!+\!\left(\varepsilon-\frac{E_0}2\right)
\frac{M^*_{FC}\!-\!M^*_L}M\left[\theta\left(
\varepsilon\!-\!\frac{E_0}2\right)
+\theta\left(\mbox{-}\varepsilon\!-\!\frac{E_0}2\right)\right].
\eeq Here $\theta(\varepsilon)$ is the step function. Note that in
order to ensure a smooth transition from the single-particle
spectrum characterized by $M^*_{FC}$ to the spectrum defined by
$M^*_{L}$ the step function is to be substituted by some smooth
function. Upon inserting Eq. (35) into Eq. (34) we can check that
inside the interval $(-E_0/2,E_0/2)$ the effective mass $M^*\simeq
M^*_{FC}$, and outside the interval $M^*\simeq M^*_{L}$. By
applying the Kramers-Kr\"{o}nig transformation to
${\mathrm{Re}}\Sigma({\bf p},\varepsilon)$, we obtain the
imaginary part of the self-energy \cite{ams} \beq \mbox{Im
}\Sigma({\bf p},\varepsilon)\sim
\varepsilon^2\frac{M^*_{FC}}{\varepsilon_F M}+
\frac{M^*_{FC}-M^*_L}M\left(\!\varepsilon\ln\!\left|
\frac{\varepsilon\!+\!E_0/2}{\varepsilon\!-\!E_0/2}\right|\!+\!
\frac{E_0}2\ln\!\left|\frac{\varepsilon^2\!-\!E^2_0/4}{E^2_0/4}
\right|\right). \eeq We see from Eq. (36) that at
$\varepsilon/E_0\ll 1$ the imaginary part is proportional to
$\varepsilon^2$, at $2\varepsilon/E_0\simeq 1$
${\mathrm{Im}}\Sigma\sim \varepsilon$, and at $E_0/\varepsilon\ll 1$
the main contribution to the imaginary part is approximately
constant. This is the behavior that gives rise to the known
peak-dip-hump structure. It is seen from Eq. (36) that when
$E_0\to 0$ the second term on the right hand side tends to zero
and the single-particle excitations become  better defined,
resembling the situation in a normal Fermi-liquid, and the
peak-dip-hump structure eventually vanishes. On the other hand,
the quasiparticle amplitude $a({\bf p})$ is given by \cite{mig}
\beq \frac1{a({\bf p})}\ =\ 1-\frac{\partial \mbox{ Re
}\Sigma({\bf p},\varepsilon)}{\partial\varepsilon}\ . \eeq
At $T\leq T_c$, as seen from Eq. (35), the quasiparticle amplitude at
the Fermi surface rises as the energy scale $E_0$ decreases.  It
follows from Eqs. (18) and (27) that $E_0\sim(x_{FC}-x)/x_{FC}$.
At $T>T_c$, it is seen from Eq. (34) that the quasiparticle
amplitude rises as the effective mass $M^*_{FC}$ decreases. As seen
from Eqs. (9) and (12), $M^*_{FC}\sim (p_f-p_i)/p_F\sim
(x_{FC}-x)/x_{FC}$. As a result, we can conclude that the amplitude
rises as the level of doping increases, while the peak-dip-hump
structure vanishes and the single-particle excitations become better
defined in highly overdoped samples. At $x>x_{FC}$, the energy scale
$E_0=0$ and the quasiparticles are normal excitations of Landau Fermi
liquid. It is worth noting that such a behavior was observed
experimentally in highly overdoped Bi2212 where the gap size is about
10~meV \cite{val1}.  Such a small size of the gap verifies that the
region occupied by the FC is small since $E_0/2\simeq \Delta_1$. Then,
recent experimental data have shown that the Landau Fermi liquid does
exist in heavily overdoped non-superconducting
La$_{1.7}$Sr$_{0.3}$CuO$_4$ \cite{nakam}.

\section {Field induced Landau Fermi liquid in Fermi liquids with FC}

Now we consider the behavior of a many-electron system with FC in
magnetic fields, assuming that the coupling constant
$\lambda_0\neq 0$ is infinitely small. As we have seen in Sec. III,
at $T=0$ the superconducting order parameter $\kappa({\bf p})$ is
finite in the FC range, while the maximum value of the
superconducting gap $\Delta_1\propto \lambda_0$ is infinitely
small. Therefore, any small magnetic field $B \neq 0$ can be
considered as a critical field and will destroy the coherence of
$\kappa({\bf p})$ and thus FC itself. To define the type of FC
rearrangement, simple energy arguments are sufficient. On one
hand, the energy gain $\Delta E_B$ due to the magnetic field $B$
is $\Delta E_B\propto B^2$ and tends to zero with $B\to 0$. On the
other hand, occupying the finite range $L_{ FC}$ in the momentum
space, the formation of FC leads to a finite gain in the ground state
energy \cite{ks}. Thus, a new ground state replacing FC should
have almost the same energy as the former one. Such a state is
given by the multiconnected Fermi spheres resembling an onion,
where the smooth quasiparticle
distribution function $n({\bf p})$ in the $L_{FC}$ range is
replaced by a multiconnected distribution $\nu({\bf p})$
\cite{asp}
\begin{equation}
\nu({\bf p})=\sum\limits_{k=1}^n\theta (p-p_{2k-1})\theta (p_{2k}-p).
\end{equation}
Here the parameters $p_i\leq p_1<p_2<\ldots <p_{2n}\leq p_f$ are
adjusted to obey the normalization condition:
\begin{equation}
\int_{p_{2k}}^{p_{2k+3}}\nu({\bf p})
\frac{d{\bf p}}{(2\pi)^3}=\int_{p_{2k}}^{p_{2k+3}}
n({\bf p})\frac{d{\bf p}}{(2\pi)^3}.
\end{equation}
For definiteness, let us consider the most interesting case of a 3D
system, while the consideration of a 2D system also goes along the same
line. We note that the idea of multiconnected Fermi spheres, with
production of new, interior segments of the Fermi surface, has
been considered already \cite{llvp,zb}. Let us assume that the
thickness of each interior block is approximately the same
$p_{2k+1}-p_{2k}\simeq \delta p$ and $\delta p$ is defined by $B$.
Then, the single-particle energy in the region $L_{FC}$ can be
fitted by
\begin{equation}
\varepsilon({\bf p}) - \mu \sim \mu\frac{\delta p}{p_F}
\left [ \sin\left(\frac{p}{\delta p}\right) + b(p) \right ].
\end{equation}
The blocks are formed since all the single particle states around
the minimum values of the fast sine function are occupied and
those around its maximum values are empty, the average occupation
being controlled by a slow function $b({\bf p})\approx\cos [\pi
n({\bf p})]$. It follows from Eq. (40) that the effective mass
$m^*$ at each internal Fermi surface is of the order of the bare
mass $M$, $m^*\sim M$.  Upon
replacing $n({\bf p})$ in Eq. (5) by $\nu({\bf p})$, defined by
Eqs. (38) and (39), and using the Simpson's rule, we
obtain that the minimum loss in the ground state energy due to
formation of the blocks is about $(\delta p)^4$. This result can
be understood by considering that the continuous FC function
$n({\bf p})$ delivers the minimum value to the energy functional
$E[n({\bf p})]$, while the approximation of $\nu({\bf p})$ by steps
of size $\delta p$ produces the minimum error of the order of
$(\delta p)^4$. On the other hand, this loss must be compensated
by the energy gain due to the magnetic field. Thus, we come to the
following relation \begin{equation} \delta p\propto \sqrt{B}.
\end{equation} When the Zeeman splitting is taken into account in
the dispersion law, Eq. (40), each of the blocks is polarized,
since their outer areas are occupied only by polarized spin-up
quasiparticles. The width of each areas in the momentum
space $\delta p_0$ is given by \begin{equation} \frac{p_F\delta
p_0}{m^*}\sim B\mu_{eff},
\end{equation}
where $\mu_{eff}\sim \mu_B$ is the effective
magnetic moment of electron.
We can consider such a polarization without altering the
previous estimates, since it follows from Eq. (41) that $\delta
p_0/\delta p\ll 1$. The total polarization $\Delta P$ is obtained
by multiplying $\delta p_0$ by the number
$N$ of the blocks, which is
proportional to $1/\delta p$, $N\sim (p_f-p_i)/\delta p$. Taking
into account Eq. (41), we obtain
\begin{equation}
\Delta P\sim m^*\frac{p_f-p_i}{\delta p}B\mu_{eff}\propto \sqrt{B},
\end{equation}
which prevails over the contribution $\sim B$ obtained
within the Landau Fermi theory. On the other hand, this
quantity can be expressed as
\begin{equation}
\Delta P\propto M^*B,
\end{equation}
where $M^*$ is the ``average'' effective mass related to the finite
density of states at the Fermi level,
\beq
M^*\sim N m^*\propto\frac{1}{\delta p}.
\eeq
We can also conclude that $M^*$ defines the specific heat.

Eq. (41) can be discussed differently, starting with a different
assumption, namely, that multiconnected
Fermi sphere can be approximated by
a single block the thickness of which is $\delta p$, while the other
blocks are located under the Fermi level being represented by the
Landau levels. Let us put $\lambda_0=0$.  Then, the energy gain due to
the magnetic field is given by $\Delta E_B\sim B^2M^*$. The energy loss
$\Delta E_{FC}$ due to rearrangement of the FC state can be estimated
using the Landau formula \cite{lan}
\begin{equation}
\Delta E_{ FC}=\int(\varepsilon({\bf p})-\mu)\delta
n({\bf p})\frac{d{\bf p}^3}{(2\pi)^3}.
\end{equation}
As we have seen above, the region occupied
by the variation $\delta
n({\bf p})$ has the length $\delta p$, while $(\varepsilon({\bf
p}) -\mu)\sim (p-p_F)p_F/M^*$. As a result, we have $\Delta E_{
FC}= \delta p^2/M^*$. Equating $\Delta E_B$ and $\Delta E_{FC}$
and taking into account that in this case $M^*\propto 1/\delta p$,
we arrive at the following relation
\begin{equation} \frac{\delta
p^2}{M^*}\propto \delta p^3\propto \frac{B^2}{\delta p}, \end{equation}
which coincides with Eq. (41). It follows from Eqs.  (43) and (44) that
the effective mass $M^*$ diverges as
\begin{equation} M^*\propto \frac{1}{\sqrt{B}}.
\end{equation}
Equation (48) shows that by applying a magnetic field $B$ the
system can be driven back into the Landau Fermi-liquid with the
effective mass $M^*(B)$ dependent on the magnetic field.
This means that the coefficients $A(B)$, $\gamma_0(B)$, and $\chi_0(B)$
in the resistivity, $\rho(T)=\rho_0+\Delta\rho$ with
$\Delta\rho=A(B)T^2$
and $A(B)\propto (M^*)^2$, specific heat, $C/T=\gamma_0(B)$, and
magnetic susceptibility depend on the effective mass in accordance
with the Landau Fermi-liquid theory. It was
demonstrated that the constancy of the well-known Kadowaki-Woods
ratio, $A/\gamma_0^2\simeq const$ \cite{kadw}, is obeyed by systems in
the highly correlated regime when the effective mass is
sufficiently large \cite{ksch}. Therefore, we are led to the
conclusion that by applying magnetic fields the system is driven
back into the Landau Fermi-liquid where the constancy of the
Kadowaki-Woods ratio is obeyed. Since the resistivity is given by
$\Delta\rho\propto (M^*)^2$ \cite{ksch}, we obtain
from Eq. (48) \beq A(B)\propto \frac{1}{B}.\eeq

At finite temperatures, the system
remains in the Landau Fermi-liquid, but
there exists a temperature $T^*(B)$, at which the polarized state
is destroyed. To calculate the function $T^*(B)$ , we observe that
the effective mass $M^*$ characterizing the single particle
spectrum cannot be changed at $T^*(B)$. In other words, at the
crossover point, we have to compare the effective mass
$M^*(T)$ defined by
$T^*(B)$, Eq. (9), and that $M^*(B)$
defined by the magnetic field $B$,
Eq. (48), $M^*(T)\sim M^*(B)$
\begin{equation} \frac{1}{M^*}\propto T^*(B)\propto \sqrt{B}.
\end{equation}
As a result, we obtain
\begin{equation}
T^*(B)\propto \sqrt{B}.
\end{equation}
At temperatures $T\geq T^*(B)$, the system comes back into the
state with $M^*$ defined by Eq. (9), and we do not observe the Landau
Fermi-liquid (LFL) behavior.
We can conclude that Eq. (51) determines the line in the $B-T$ phase
diagram which separates the region of the $B$ dependent
effective mass from the
region of the $T$ dependent effective mass, see also Sec. VII.
At the temperature $T^*(B)$, there occurs a
crossover from the $T^2$ dependence
of the resistivity to the $T$ dependence.
It follows from Eq. (51), that a heavy
fermion system at some temperature $T$ can be driven back into the
Landau Fermi-liquid by applying a strong enough magnetic field
$B\geq B_{cr}\propto (T^*(B))^2$.  We can also conclude, that at
finite temperature $T$, the effective mass of a heavy fermion
system is relatively field-independent at magnetic fields $B\leq
B_{cr}$ and show a more pronounced metallic behavior at $B\geq
B_{cr}$, since the effective mass decreases (see Eq. (48)). The
same behavior of the effective mass can be observed in the
Shubnikov-de Haas oscillation measurements. We note that our
consideration is valid for temperatures $T\ll T_f$.
From Eqs. (50) and (51) we obtain a unique possibility to
control the essence of the strongly correlated liquid by weak
magnetic fields which induce the change of the non-Fermi liquid
(NFL) behavior to the LFL liquid behavior.

Now we can consider the nature of the field-induced quantum
critical point in YbRh$_2$Si$_2$. The properties of this
antiferromagnetic (AF) heavy fermion metal with the ordering
Ne\`{e}l temperature $T_N=70$ mK were recently investigated in
Refs.\cite{gen,ishida}. In the AF state, this metal shows LFL
behavior. As soon as the weak AF order is suppressed
either by a tiny volume expansion or by temperature, pronounced
deviations from the LFL behavior are observed. The
experimental facts show that the spin density wave picture failed
when considering the data obtained \cite{gen,ishida,col1}. We
assume that the electron density in YbRh$_2$Si$_2$ is close to the
critical value $(x_{FC}-x)/x_{FC}\ll1$ \cite{shag1,shag2}, so that the
state with FC can be easily suppressed by weak magnetic fields or
by the AF state. In the AF state,
the effective mass is finite and the electron system of
YbRh$_2$Si$_2$ possesses the LFL behavior. When the AF
state is suppressed at $T>T_N$ the system comes back into NFL.
By tuning $T_N\to 0$ at a critical field $B=B_{c0}$, the itinerant
AF order is suppressed and replaced by spin fluctuations
\cite{ishida}. Thus, we can expect the absence of any long-ranged
magnetic order in this state, and the situation corresponds to a
paramagnetic system with strong correlations without the field, $B=0$.
As a result, the FC state is restored and we can observe NFL
behavior at any temperatures in accordance with experimental facts
\cite{gen}.  As soon as an excessive magnetic field $B>B_{c0}$ is
applied, the system is driven back into LFL. To
describe the behavior of the effective mass, we can use Eq. (48)
substituting $B$ by $B-B_{c0}$
\begin{equation}
M^*\propto \frac{1}{\sqrt{B-B_{c0}}}.
\end{equation}
Equation (52) demonstrates the $1/\sqrt{B-B_{c0}}$ divergence of the
effective mass, and therefore the coefficients $\gamma_0(B)$ and
$\chi_{0}(B)$ should have the same behavior. Meanwhile the coefficient
$A(B)$ diverges as $1/(B-B_{c0})$, being proportional to $(M^*)^2$
\cite{ksch}, and thus preserving the Kadowaki-Woods ratio, in
agreement with the experimental finding \cite{gen}.
To construct a $B-T$
phase diagram for YbRh$_2$Si$_2$ we use the same replacement $B\to
B-B_{c0}$ in Eq. (51) so that
\begin{equation} T^*(B)\simeq c\sqrt{B-B_{c0}},
\end{equation}
where $c$ is a constant.

The $B-T$ phase diagram given by Eq. (53) is in good quantitative
agreement with the experimental data \cite{gen}.
This phase diagram has to have a strong impact on the
magnetoresistance of YbRh$_2$Si$_2$ and other metals with heavy
fermions having similar $B-T$ phase diagrams, see Sect. VII. We note
that our consideration is valid at temperatures $T \ll T_f$. The
experimental phase diagram shows that the behavior $T^*\propto
\sqrt{B-B_{c0}}$ is observed up to $150$ mK \cite{gen} and allows
us to estimate the magnitude of $T_f$, which can reach at least
$1$ K in this system. We can conclude that a new type of the
quantum critical point observed in a heavy-fermion metal
YbRh$_2$Si$_2$ can be identified as FCQPT with
the order parameter $\kappa({\bf p})$ and with the gap $\Delta_1$
being infinitely small \cite{pogsh}.

It was reported recently that in the normal state obtained by
applying a magnetic field greater than the upper critical filed
$B_c$ suppressing the superconductivity, in a hole-doped cuprates at
overdoped concentration (Tl$_2$Ba$_2$CuO$_{6+\delta}$) \cite{cyr} and
at optimal doping concentration (Bi$_2$Sr$_2$CuO$_{6+\delta}$)
\cite{cyr1}, there are no any sizable violations of the Wiedemann-Franz
(WF) law.  In the electron-doped copper oxide superconductor
Pr$_{0.91}$LaCe$_{0.09}$Cu0$_{4-y}$ ($T_c$=24 K) when
superconductivity is removed by a strong magnetic field,
it was found that the spin-lattice relaxation rate $1/T_1$
follows the $T_1 T=constant$ relation, known as Korringa law
\cite{korr}, down to temperature of $T=0.2$ K \cite{zheng}. At
elevated temperatures and applied magnetic fields of 15.3 T
perpendicular to the CuO$_2$ plane, $1/T_1T$ as a function of $T$ is
a constant below $T=55$ K.
At $300$ K $>T>50$ K, $1/T_1T$  decreases with
increasing $T$ \cite{zheng}. Recent measurements for strongly
overdoped non-superconducting La$_{1.7}$Sr$_{0.3}$CuO$_4$ have shown
that the resistivity $\rho$ exhibits $T^2$ behavior,
$\rho=\rho_0+\Delta\rho$ with $\Delta\rho=AT^2$, and the WF law is
verified to hold perfectly \cite{nakam}. Since the validity of the WF
law and of the Korringa law are a robust signature of LFL, these
experimental facts demonstrate that the observed elementary
excitations cannot be distinguished from the Landau quasiparticles.

If $\lambda_0$ is finite, the critical
field is also finite, and Eq. (50) is valid at $B>B_c$. In that
case, the effective mass $M^*_{FC}(B)$ given by Eq.  (17) is finite,
and the system is driven back to LFL and has the LFL behavior induced
by the magnetic field.  At a constant magnetic field, the low energy
elementary excitations are characterized by $M^*_{FC}(B)$ and cannot be
distinguished from Landau quasiparticles.  As a result, at $T\to 0$,
the WF law is held in accordance with experimental facts
\cite{cyr,cyr1}.  On the hand, in contrast to the LFL theory, the
effective mass  $M^*_{FC}(B)$ depends on the magnetic field.

Equation (50) shows that by applying a magnetic field $B>B_c$ the
system can be driven back into LFL with the effective mass
$M^*_{FC}(B)$ which is finite and independent of the temperature.
This means that the low temperature properties depend on the
effective mass in accordance with the LFL theory. In particular,
the resistivity $\rho(T)$ as a function of the temperature behaves
as $\rho(T)=\rho_0+\Delta\rho(T)$ with $\Delta\rho(T)=AT^2$, and the
factor $A\propto (M^*_{FC}(B))^2$.  At finite temperatures, the system
persists to be LFL, but there is the crossover from the LFL
behavior to the non-Fermi liquid behavior at temperature
$T^*(B)\propto \sqrt{B}$, see Eq. (51). At $T>T^*(B)$, the effective
mass starts to depend on the temperature $M_{FC}\propto 1/T$, and the
resistivity possesses the non-Fermi liquid behavior with a
substantial linear term, $\Delta\rho(T)=aT+bT^2$ \cite{ms,pogsh}.
Such a behavior of the resistivity was observed in the
cuprate superconductor Tl$_2$Ba$_2$CuO$_{6+\delta}$
($T_c<$ 15 K) \cite{mac}. At B=10 T, $\Delta\rho(T)$ is a
linear function of the temperature
between 120 mK and 1.2 K, whereas at B=18 T, the temperature
dependence of the resistivity is consistent with
$\rho(T)=\rho_0+AT^2$ over the same temperature range \cite{mac}.

In LFL, the nuclear spin-lattice relaxation rate $1/T_1$
is determined by the quasiparticles near the Fermi level whose
population is proportional to $M^* T$, so that $1/T_1T\propto M^*$
is a constant \cite{zheng,korr}. As it was shown, when the
superconducting state is removed by the application of a magnetic
field, the underlying ground state can be seen as the field induced
LFL with effective mass depending on the magnetic field.  As a
result, the rate $1/T_1$ follows the $T_1T=constant$ relation, that
is the Korringa law is held.  Unlike the behavior of LFL, as it
follows from Eq. (50), $1/T_1T\propto M^*_{FC}(B)$ decreases with
increasing the magnetic field at $T<T^*(B)$.  While, at $T>T^*(B)$,
we observe that $1/T_1T$ is a decreasing function of the temperature,
$1/T_1T\propto M^*_{FC}\propto 1/T$.  These observations are in a
good agreement with the experimental facts \cite{zheng}.  Since
$T^*(B)$ is an increasing function of the magnetic field, the
Korringa law retains its validity to higher temperatures at elevated
magnetic fields. We can also conclude, that at temperature $T_0\leq
T^*(B_0)$ and elevated magnetic fields $B>B_0$, the system shows a
more pronounced metallic behavior since the effective mass decreases
with increasing $B$, see Eq. (50). Such a behavior of the effective
mass can be observed in the de Haas van Alphen-Shubnikov studies,
$1/T_1T$ measurements, and the resistivity measurements \cite{aspla}.
These experiments can shed light on the physics of high-$T_c$ metals
and reveal relationships between high-$T_c$ metals and heavy-electron
metals.

\section{ Appearance of FCQPT in different Fermi liquids}

It is widely believed that unusual properties of the strongly
correlated liquids observed in the high-temperature
superconductors, heavy-fermion metals, 2D $^3$He and etc., are
determined by quantum phase transitions.
Therefore, immediate experimental studies of relevant quantum phase
transitions and of their quantum critical points
are of crucial importance for understanding the physics of the
high-temperature superconductivity and strongly correlated systems.  In
case of the high-temperature superconductors, these studies are
difficult to carry out, because all the corresponding area is occupied
by the superconductivity. On the other hand, recent experimental data
on different Fermi liquids in the highly correlated regime at the
critical point and above the point
can help to illuminate both the nature of this point
and the control parameter by which this point is driven. Experimental
facts on strongly interacting high-density two dimensional (2D) $^3$He
\cite{mor,cas} show that the effective mass diverges when the density
at which 2D $^3$He liquid begins to solidify is approached
\cite{cas}. Then, a sharp increase of the effective mass
in a metallic 2D electron system was observed, when the
density tends to the critical density of the metal-insulator
transition point, which occurs at sufficiently low densities
\cite{skdk}. Note, that
there is no ferromagnetic instability in both Fermi systems and the
relevant Landau amplitude $F^a_0>-1$ \cite{skdk,cas}, in accordance
with the almost localized fermion model \cite{pfw}.

Now we consider the divergence of the effective mass in 2D and 3D
Fermi liquids at $T=0$, when the density $x$ approaches FCQPT from
the side of normal Landau Fermi liquid, that is from the disordered
phase.  First, we calculate the divergence of $M^*$ as a function of
the difference $(x_{FC}-x)$ in case of 2D $^3$He. For this purpose we
use the equation for $M^*$ obtained in \cite{ksz}, where the divergence
of the effective mass $M^*$ due to the onset of FC in different Fermi
liquids including $^3$He was predicted. At $x\to x_{FC}$, the effective
mass $M^*$ can be approximated as
\beq\frac{1}{M^{*}}\simeq\frac{1}{M}+\frac{1}{4\pi^{2}}
\int\limits_{-1}^{1}\int\limits_0^{g_0}
\frac{v(q(y))}{\left[1-R(q(y),\omega=0,g)
\chi_0(q(y),\omega=0)\right]^{2}}\frac{ydydg}{\sqrt{1-y^{2}}}.
\eeq Here we adopt the notation $p_F\sqrt{2(1-y)}=q(y)$ with
$q(y)$ being the transferred momentum, $M$ is the bare mass,
$\omega$ is the frequency, $v(q)$ is the bare interaction, and the
integral is taken over the coupling constant $g$ from zero to its
real value $g_0$. In Eq.  (54), both $\chi_0(q,\omega)$ and
$R(q,\omega)$, being the linear response function of
a noninteracting Fermi liquid and the effective interaction
respectively, define the linear response function of the system in
question \beq \chi(q,\omega,g)=\frac{\chi_0(q,\omega)}
{1-R(q,\omega,g)\chi_0(q,\omega)}. \eeq
In the vicinity of the charge
density wave instability, occurring at the density $x_{cdw}$, the
singular part of the function $\chi^{-1}$ on the disordered side
is of the well-known form (see  e.g.  \cite{vn})
\beq\chi^{-1}(q,\omega,g)\propto
(x_{cdw}-x)+(q-q_c)^2+(g_0-g),\eeq where $q_c\sim 2p_F$ is the
wavenumber of the charge density wave order. Upon substituting Eq.
(56) into Eq. (54) and integrating, the equation for the
effective mass $M^*$ can be cast into the following form
\beq\frac{1}{M^*}= \frac{1}{M}-\frac{C}{\sqrt{x_{cdw}-x}},\eeq
with $C$ being some positive constant. It is seen from Eq. (57)
that $M^*$ diverges at some point $x_{FC}$ referred to as the
critical point, at which FCQPT occurs as a function of the
difference $(x_{FC}-x)$ \cite{shag1}
\beq M^*\propto \frac{1}{x_{FC}-x}.\eeq
It follows from the derivation of Eqs. (57) and (58) that
their forms are
independent of the bare interaction $v(q)$. Therefore
both of these equations are also applicable to 2D electron liquid
or to another Fermi liquid. It is also seen from Eqs. (57) and
(58) that FCQPT precedes the formation of charge-density waves. As
a consequence of this, the effective mass diverges at high
densities in case of 2D $^3$He, and at low densities in
case of 2D electron systems, in accordance with experimental facts
\cite{skdk,cas}. Note, that in both cases the difference
$(x_{FC}-x)$ has to be positive, because $x_{FC}$ represents the
solution  of Eq. (57). Thus, in considering the many-electron systems
we have to replace $(x_{FC}-x)$ by $(x-x_{FC})$.  In case of a 3D
system, at $x\to x_{FC}$, the effective mass is given by \cite{ksz}
\beq\frac{1}{M^{*}}\simeq\frac{1}{M}+\frac{p_F}{4\pi^{2}}
\int\limits_{- 1}^{1}\int\limits_0^{g_0}
\frac{v(q(y))ydydg}{\left[1-R(q(y),\omega=0,g)
\chi_0(q(y),\omega=0)\right]^{2}}. \eeq A comparison of Eq. (54)
and Eq. (59) shows that there is no fundamental difference between
these equations, and along the same way we again arrive at Eqs.
(57) and (58). The only difference between 2D electron systems and
3D ones is that in the latter FCQPT occurs at densities which are well below
those corresponding to 2D systems. For bulk $^3$He, FCQPT
cannot probably take place since it is absorbed by the first order
solidification \cite{cas}.

Now we address the problem of the fermion condensation in dilute
Fermi gases and in a low density neutron matter. We consider an
infinitely extended system composed of Fermi particles, or atoms,
interacting by an artificially constructed potential with the
desirable scattering length $a$. These objects may be viewed as
trapped Fermi gases, which are systems composed of Fermi atoms
interacting by a potential with almost any desirable scattering
length, similarly to that done for the trapped Bose gases, see
e.g. \cite{nat}. If $a$ is negative the system becomes unstable at
densities $x \sim |a|^{-3}$, provided the scattering length is the
dominant parameter of the problem. That means that $|a|$ is much
bigger than the radius of the interaction or any other relevant
parameter of the system. The compressibility $K(x)$ vanishes at
the density $x_{c1}\sim |a|^{-3}$, making the system completely
unstable \cite{ams5}. Expressing the linear response function in
terms of the compressibility \cite{lanl1},
\begin{equation}
\chi (q\rightarrow 0,i\omega \rightarrow 0)
=-\left(\frac{d^{2}E}{dx^{2}}
\right)^{-1},
\end{equation}
we obtain that the linear response function has a pole at the
origin of coordinates, $q\simeq 0,\,\omega\simeq 0$, at the same point
$x_{c1}$. To find the behavior of the effective mass $M^*$ as a
function of the density, we substitute Eq. (56) into Eq. (59)
taking into account that $x_{cdw}=x_{c1}$ and $q_c/p_F\ll 1$ due
to Eq. (60). At low momenta $q/p_F\sim 1$, the potential $v(q)$ is
attractive because the scattering length is the dominant parameter
and negative. Therefore, the integral on the right hand side of
Eq. (59) is negative and diverges at $x\to x_{1c}$. The above
considerations can also be applied to the
clarification of the fact that
the effective mass $M^*$ is again given by Eq. (58) with
$x_{FC}<x_{c1}$. Note that the superfluid correlation cannot stop
the system from squeezing, since their contribution to the ground
state energy is negative. After all, the superfluid correlations
can be considered as additional degrees of freedom, which can
therefore only decrease the energy.  We conclude that FCQPT can be
observed in traps by measuring the density of states at the Fermi
level, which becomes extremely large as $x\to x_{FC}$. Note that at
these densities the system remains stable because $x_{FC}<x_{c1}$.
It seems quite probable that the neutron-neutron scattering length
($a\simeq -20$ fm) is sufficiently large to be the dominant
parameter and  to permit the neutron matter to have an equilibrium
energy, density, and the singular point $x_{c1}$, at which the
compressibility vanishes \cite{ams4}. Therefore, we can expect
that FCQPT takes place in a low density neutron matter leading to
stabilization of the matter by lowering its ground state energy. A
more detailed analysis of this possibility will be published
elsewhere.

A few remarks are in order.
We have seen that above the critical point $x_{FC}$ the effective mass
$M^*$ is finite and, therefore, the
system exhibits the Landau Fermi liquid behavior.
If $|x-x_{FC}|/x_{FC}\ll 1$ the behavior can be viewed as
a highly correlated one because the effective mass,
being given by Eq.
(58), strongly depends on the density,
temperature and magnetic fields and is very large, see Sec.  VII.
Beyond this region, the effective mass is approximately constant and
the system becomes a normal Landau Fermi liquid.  We can expect to
observe such a highly correlated electron (or hole) liquid in heavily
overdoped high-T$_c$ compounds which are located beyond the
superconducting dome.  We recall that beyond the FCQPT point the
superconducting gap $\Delta_1$ can be very small or even absent, see
Eq. (28).  Indeed, recent experimental data have shown that this
liquid exists in heavily overdoped non-superconducting
La$_{1.7}$Sr$_{0.3}$CuO$_4$ \cite{nakam}.

\section{Field-induced
Landau Fermi liquid in highly correlated Fermi liquids}

When a Fermi system approaches
FCQPT from the disordered phase
it remains the Landau Fermi liquid with the effective mass $M^*$
strongly depending on the deviation of the density $\Delta
x=|x_{FC}-x|$, temperature and a magnetic field $B$ provided that
$|x_{FC}-x|/x_{FC}\ll 1$ and $T\leq T^*(x)$ \cite{shag1,shag2}.  This
state of the system, with $M^*$ strongly depending on $T$, $\Delta x$
and $B$, resembles the strongly correlated liquid. In contrast to a
strongly correlated liquid, there is no the energy scale $E_0$ and the
system under consideration is the Landau Fermi liquid at sufficiently
low temperatures $T\leq T^*(x)$ with the effective mass
$M^*\propto1/\Delta x$. Therefore this liquid can be called a highly
correlated liquid.  Obviously, a highly correlated liquid is to have
uncommon properties.

In this Section, we study the behavior of a highly correlated electron
liquid in magnetic fields. We show that at $T\geq T^*(x)$
the effective mass starts to depend on the temperature,
$M^*\propto T^{-1/2}$. This $T^{-1/2}$ dependence of the effective mass
at elevated  temperatures leads to the non-Fermi liquid behavior of
the resistivity,  $\rho(T)\sim \rho_0+aT+bT^{3/2}$. The
application of magnetic field $B$ restores the common $T^2$ behavior of
the resistivity, $\rho\simeq \rho_0+AT^2$ with  $A\propto (M^*)^2$.
Both the effective mass and coefficient $A$ depend on the magnetic field,
$M^*(B)\propto B^{-2/3}$ and $A\propto B^{-4/3}$ being approximately
independent of the temperature at $T\leq T^*(B)\propto B^{4/3}$.
At $T\geq T^*(B)$, the $T^{-1/2}$ dependence of the effective mass is
re-established. We demonstrate that this $B-T$ phase diagram has a
strong impact on the magnetoresistance (MR) of the highly correlated
electron liquid. The MR as a function of the temperature exhibits a
transition from the negative values of MR at $T\to 0$ to the positive
values at $T>T^*(B)\propto B^{4/3}$. Thus, at $T\geq T^*(B)$,
MR as the function of the temperature possesses a node at
$T\propto B^{4/3}$. Such a behavior is
of general form and takes place in both 3D
highly correlated systems and 2D ones.

It follows from Eq. (58) that effective mass
is finite provided that $|x-x_{FC}|\equiv \Delta x>0$. Therefore,
the system represents the Landau Fermi liquid.
In case of electronic systems the Wiedemann-Franz law is held at
$T\to 0$, and Kadowaki-Woods ratio is preserved.
Beyond the region $|x-x_{FC}|/x_{FC}\ll 1$, the
effective mass is approximately constant
and the system becomes  conventional Landau Fermi liquid.
On the other hand, $M^*$ diverges as the
density $x$ tends to the critical point of FCQPT. As a result,
the effective mass strongly depends on such quantities as the
temperature, pressure, magnetic field given that they exceed
their critical values.
For example, when
$T$ exceeds some temperature $T^*(x)$, Eq. (58) is no longer valid,
and $M^*$ depends on the temperature as well.
To evaluate this dependence, we calculate the deviation
$\Delta x(T)$ generated by $T$.
The temperature smoothing out the Fermi function $\theta(p_F-p)$
at $p_F$ induces the variation $p_F \Delta p/M^*(x)\sim T$, and
$\Delta x(T)/x_{FC}\sim M^*(x)T/p_F^2$, with $p_F$ is the
Fermi momentum and $M$ is the bare electron mass.
The deviation $\Delta x$ can be
expressed in  terms of $M^*(x)$ using Eq. (58),
$\Delta x/x_{FC}\sim M/M^*(x)$.
Comparing these deviations,
we find that at $T\geq T^*(x)$  the effective mass depends
noticeably on the temperature, and
the equation for $T^*(x)$ becomes
\beq T^*(x)\sim p_F^2\frac{M}{(M^*(x))^2}\sim \varepsilon_F(x)
\left(\frac{M}{M^*(x)}\right)^2.\eeq
Here $\varepsilon_F(x)$ is the Fermi energy of noninteracting electrons
with mass $M$. It follows from Eq. (61) that $M^*$ is always finite
at temperatures $T>0$. We can consider $T^*(x)$ as the energy scale
$e_0(x)\simeq T^*(x)$. This scale defines the area
$(\mu -e_0(x))$ in the single
particle spectrum where $M^*$ is approximately constant, being given by
$M^*=d\varepsilon(p)/dp$ \cite{lan}. According  to Eqs. (58) and (61) it
is easily verified that $e_0(x)$ can be written in the form
\beq e_0(x)\sim \varepsilon_F
\left(\frac{x-x_{FC}}{x_{FC}}\right)^2. \eeq
At $T\ll e_0(x)$ and above the critical point the effective mass
$M^*(x)$ is finite, the energy scale $E_0$ given by Eq. (62) vanishes
and the system exhibits the LFL behavior.
At temperatures $T \geq e_0(x)$ the effective mass $M^*$ starts to
depend on the temperature and the NFL behavior is observed.
Thus, at $|x-x_{FC}|/x_{FC}\ll 1$ the system can be considered as a
highly correlated one: at $T\ll e_0(x)$, the system is LFL, while
at temperatures $T \geq e_0(x)$,  the system possesses the NFL behavior.

At $T\geq T^*(x)$, the main contribution to
$\Delta x$ comes from the temperature, therefore
\beq
M^*\sim M\frac{x_{FC}}{\Delta x(T)}\sim M\frac{\varepsilon_F}{M^*T}.\eeq
As a result, we obtain
\beq
M^*(T)\sim M\left(\frac{\varepsilon_F}{T}\right)^{1/2}.\eeq
Equation (64) allows us to evaluate the resistivity as a function of $T$.
There are two terms contributing to the resistivity. Taking into account
that $A\sim (M^*)^2$ and Eq. (64), we obtain the first term
$\rho_1(T)\sim T$. The second term $\rho_2(T)$ is related to
the quasiparticle width
$\gamma$. When $M/M^*\ll 1$, the width
$\gamma\propto (M^*)^3 T^2/\epsilon(M^*)\propto T^{3/2}$,
with $\epsilon(M^*)\propto (M^*)^2$ is the dielectric constant
\cite{ars,ksch}. Combining both of the contributions, we find that
the resistivity is given by
\beq \rho(T)-\rho_0\sim aT+bT^{3/2}.\eeq
Here $a$ and $b$ are constants.
Thus, it turns out that at low temperatures, $T<T^*(x)$, the
resistivity $\rho(T)-\rho_0\sim AT^2$. At higher temperatures, the
effective mass depends on the temperature and
the main contribution comes from the first term on the right hand side
of Eq. (65). While $\rho(T)-\rho_0$ follows the $T^{3/2}$ dependence
at elevated temperatures.

In the same way as Eq. (64) was derived, we can obtain the equation
determining $M^*(B)$ \cite{shag2}. The application of magnetic field
$B$ leads to a weakly polarized state, or Zeeman splitting, when some
levels at the Fermi level are
occupied by spin-up polarized quasiparticles. The width
$\delta p=p_{F1}-p_{F2}$ of the area in
the momentum space occupied by these quasiparticles is of the order
$$ \frac{p_F\delta p}{M^*}\sim B\mu_{eff}.$$
Here $\mu_{eff}\sim \mu_B$ is the electron magnetic
effective moment, $p_{F1}$ is the Fermi momentum of the spin-up electrons,
and $p_{F2}$ is the Fermi momentum of the spin-down electrons. As a result,
the Zeeman splitting leads to the change $\Delta x$ in the density $x$
$$ \frac{\Delta x}{x_{FC}}\sim \frac{\delta p^2}{p_F^2}. $$
We assume that $\Delta x/x_{FC}\ll 1$.
Now it follows that
\beq M^*(B)\sim M\left(\frac{\varepsilon_F}
{B\mu_{eff}}\right)^{2/3}. \eeq
We note that $M^*$ is determined by Eq. (66) as long as
$M^*(B)\leq M^*(x)$, otherwise we have to use Eq. (58).
It follows from Eq. (66) that the application of a magnetic field reduces
the effective mass.
Note, that if there exists an itinerant magnetic order in the system
which is suppressed by magnetic field $B=B_{c0}$, Eq. (66) has to
be replaced by the equation \cite{pogsh}, see also Sec. V,
\beq M^*(B)\propto \left(\frac{1}
{B-B_{c0}}\right)^{2/3}.\eeq
The coefficient $A(B)\propto (M^*(B))^2$ diverges as
\beq A(B)\propto \left(\frac{1}
{B-B_{c0}}\right)^{4/3}.\eeq
At elevated temperature, there is a temperature $T^*(B)$ at which
$M^*(B)\simeq M^*(T)$. Comparing Eq. (64) and Eq. (67), we see that
$T^*(B)$ is given by
\beq T^*(B)\propto (B-B_{c0})^{4/3}. \eeq
At $T\geq T^*(x)$, Eq. (69) determines the line in the $B-T$ phase
diagram which separates the region of the
$B$ dependent effective mass from the
region of the $T$ dependent effective mass.
At the temperature $T^*(B)$,  a
crossover from the $T^2$ dependence
of the resistivity to the $T$ dependence occurs: at $T<T^*(B)$,
the effective mass is given by Eq. (67), and at $T>T^*(B)$
$M^*$ is given by Eq. (64).

Using the $B-T$ phase diagram just presented,
we consider the behavior of MR
\beq \rho_{mr}(B,T)=\frac{\rho(B,T)-\rho(0,T)}{\rho(0,T)},\eeq
as a function of magnetic
field $B$ and $T$. Here $\rho(B,T)$ is the
resistivity measured at the magnetic field
$B$ and temperature $T$. We assume that the contribution
$\Delta\rho_{mr}(B)$
coming from the magnetic field $B$ can be treated within
the low field approximation and given by the
well-known Kohler's rule,
\beq\Delta\rho_{mr}(B)\sim B^2\rho(0,\Theta_D)/\rho(0,T),\eeq
with $\Theta_D$ is the Debye temperature. Note, that the low
field approximation implies that
$\Delta\rho_{mr}(B)\ll \rho(0,T)\equiv\rho(T)$.
Substituting Eq. (71)
into Eq. (70), we find that
\beq \rho_{mr}(B,T)\sim
\frac{c(M^*(B,T))^2T^2+\Delta\rho_{mr}(B)-c(M^*(0,T))^2T^2}
{\rho(0,T)}.\eeq
Here $M^*(B,T)$ denotes the effective mass $M^*$ which now depends
on both the magnetic field and the temperature, and $c$ is a constant.

Consider MR given by Eq. (72) as a function
of $B$ at some temperature $T=T_0$. At low temperatures
$T_0\leq T^*(x)$, the system behaves as common Landau Fermi liquid,
and MR is an increasing function of $B$. When the
temperature $T_0$ is sufficiently high, $T^*(B)<T_0$, and
the magnetic field is  small, $M^*(B,T)$ is given by Eq. (64). Therefore,
the difference $\Delta M^*=|M^*(B,T)-M^*(0,T)|$ is small and the main
contribution is given by $\Delta\rho_{mr}(B)$.
As a result, MR is an increasing function of $B$.
At elevated $B$, the difference
$\Delta M^*$ becomes a decreasing function of $B$, and MR as the function
of $B$ reaches its maximum value at $T^*(B)\sim T_0$. In accordance
with Eq. (69), $T^*(B)$ determines the crossover from $T^2$ dependence
of the resistivity to the $T$ dependence.
Differentiating the function $\rho_{mr}(B,T)$ given by Eq.
(72) with respect
to $B$, one can verify that the derivative
is negative at sufficiently large values
of the magnetic field when $T^*(B)\simeq T_0$.
Thus, we are led to the conclusion that the crossover manifests itself as
the maximum of MR as the function of $B$.

We now consider MR as a function of $T$ at some $B_0$.
At $T\leq T^*(x)$, we have a normal Landau liquid.
At low temperatures $T^*(x)\leq T\ll T^*(B_0)$, it follows
from Eqs. (64) and (67) that $M^*(B_0)/M^*(T)\ll 1$, and
MR is determined by the resistivity $\rho(0,T)$.
Note, that $B_0$ has to be comparatively high to ensure the inequality,
$T^*(x)\leq T\ll T^*(B_0)$.
As a result,
MR tends to $-1$, $\rho_{mr}(B_0,T\to0) \simeq -1$.
Differentiating the function
$\rho_{mr}(B_0,T)$
with respect to $B_0$ we can check that its slope becomes steeper
as $B_0$ is decreased, being proportional $\propto (B_0-B_{c0})^{-7/3}$.
At $T=T_1\sim T^*(B_0)$, MR possesses a node because at this
point the effective mass $M^*(B_0)\simeq M^*(T)$,
and $\rho(B_0,T)\simeq \rho(0,T)$.
Again, we can conclude that the crossover from the $T^2$
resistivity to the $T$ resistivity, which occurs at $T\sim T^*(B_0)$,
manifests itself in
the transition from negative MR to positive MR.
At $T>T^*(B_0)$,
the main contribution
to MR comes from $\Delta\rho_{mr}(B_0)$, and MR reaches its maximum value.
Upon using Eq. (71) and taking into account that at this point
$T$ has to be determined by Eq. (69), $T\propto (B_0-B_{c0})^{4/3}$,
we obtain that the maximum value
$\rho^m_{mr}(B_0)$ of MR is
$\rho^m_{mr}(B_0)\propto (B-B_{c0})^{-2/3}$.
Thus, the maximum value is a decreasing function of $B_0$.
At $T^*(B_0)\ll T$, MR is a decreasing function of the temperature, and
at elevated temperatures MR eventually
vanishes since $\Delta\rho_{mr}(B_0)/\rho(T)\ll 1$.

The recent paper \cite{pag} reports on measurements of
the resistivity of CeCoIn$_5$ in
a magnetic field.
With increasing field, the resistivity evolves from the
$T$ temperature dependence to the $T^2$ dependence, while the field
dependence of $A(B)\sim (M^*(B))^2$ displays the critical behavior best
fitted by the function,
$A(B)\propto (B-B_{c0})^{-\alpha}$, with $\alpha\simeq 1.37$ \cite{pag}.
All these facts are in a good agreement
with the $B-T$ phase diagram given by Eq. (69). The critical behavior
displaying $\alpha=4/3$ \cite{shag2} and described
by Eq. (68) is also in a good agreement with the data.
A transition from negative MR to positive MR with increasing
$T$ was also observed \cite{pag}. We believe that an additional
analysis of the data \cite{pag} can reveal that the crossover
from $T^2$ dependence of the resistivity to the $T$ dependence
occurs at $T\propto (B-B_{c0})^{4/3}$. As well, this analysis
could reveal supplementary peculiarities of MR.
While the behavior of
the heavy fermion metal CeCoIn$_5$ in magnetic fields displayed
in Ref. \cite{pag} can be identified
as the highly correlated
behavior of a Landau Fermi liquid approaching FCQPT from the disordered
phase \cite{shag2}. Recent measurements on the heavy metal compound
CeRu$_2$Si$_2$ carried out at microkelvin temperatures down to
170 $\mu$K shows that the critical field $B_{c0}$ can be at least as
small as 0.02 mT \cite{taka}. Our analysis of the behavior of
CeRu$_2$Si$_2$ based on the experimental evidences \cite{taka} shows
that the electron system of this metal displays the highly correlated
behavior similar to the behavior of CeCoIn$_5$. The results will be
published elsewhere.

\section{Summary and conclusion}

We have discussed the appearance of the fermion condensation,
which can be compared  to the Bose-Einstein condensation. A number
of experimental evidences have been presented that are supportive
to the idea of the existence of FC in different liquids. We have
demonstrated also that experimental facts collected in different
materials,  belonging  to the high-$T_c$ superconductors, heavy
fermion metals and strongly correlated 2D structures, can be
explained within the framework of the theory based on FCQPT.

We have shown that the  appearance of FC is a quantum phase
transition, that separates the regions of normal and strongly
correlated liquids. Beyond the FCQPT point the
quasiparticle system is divided into two subsystems, one
containing normal quasiparticles, the other being occupied by
fermion condensate localized at the Fermi level. In the
superconducting state the quasiparticle
dispersion $\varepsilon({\bf p})$ in systems with FC can be
represented by two straight lines, characterized by effective masses
$M^*_{FC}$ and $M^*_L$, and intersecting near the binding energy $E_0$
which is of the order of the superconducting gap.  The same
quasiparticle picture and the energy scale $E_0$ persist in the normal
state. We have demonstrated that fermion systems with FC have features
of a "quantum protectorate"  and shown that the theory of high
temperature superconductivity, based on the fermion condensation
quantum phase transition and on the conventional theory of
superconductivity, permits the description of high values of $T_c$ and
of the maximum value of the gap $\Delta_1$, which may be as big as
$\Delta_1\sim 0.1\varepsilon_F$ or even larger.  We have also traced
the transition from conventional superconductors to high-$T_c$ ones.
We have shown by a simple, although self-consistent analysis that the
general features of the shape of the critical temperature $T_c(x)$ as a
function of the density $x$ of the mobile carriers in the high-$T_c$
compounds can be understood within the framework of the theory.

We have demonstrated that strongly correlated electron
liquids with FC, which exhibit strong deviations from the Landau
Fermi liquid behavior, can be driven into the Landau Fermi liquid by
applying a magnetic field $(B-B{c0})$ at low temperatures. A
re-entrance into the strongly correlated regime is observed if the
magnetic field $(B-B{c0})$ decreases to zero, while the effective mass
$M^*$ diverges as $M^*\propto 1/\sqrt{B-B{c0}}$. The regime is restored
at some temperature $T^*\propto\sqrt{B-B{c0}}$.
This $B-T$ phase diagram is in good quantitative
agreement with the experimental data and allow us to  conclude that
a new type of the quantum critical point observed in a heavy-fermion
metal YbRh$_2$Si$_2$ can be identified as FCQPT with the order
parameter $\kappa({\bf p})$ and with the gap $\Delta_1$ being
infinitely small. This behavior is of
a general form and takes place in both three dimensional and two
dimensional strongly correlated systems, and demonstrates the
possibility to control the essence of strongly correlated electron
liquids including heavy-fermion metals by  magnetic fields.

The appearance of FCQPT in 2D strongly correlated structures,
in trapped Fermi gases and in a low density neutron matter has been
considered. We have provided an explanation of the experimental
data on the divergence of the effective mass in a 2D electron liquid
and in 2D $^3$He, as well as shown that above the critical point the
system exhibits the Landau Fermi liquid behavior. We expect
that FCQPT takes place in trapped Fermi gases and in
a low density neutron matter leading to
stabilization of the matter by lowering its ground state energy.
At $|x-x_{FC}|/x_{FC}\ll 1$, the behavior of a Fermi liquid
approaching FCQPT form the disordered phase can be viewed as a highly
correlated one because the effective mass is very large and strongly
depends on the density, temperature and magnetic fields. Beyond this
region, the effective mass is approximately constant and the system
becomes a normal Landau Fermi liquid.

The behavior in magnetic fields of a highly correlated electron
liquid approaching FCQPT from the disordered phase
has been considered. We have shown that at sufficiently high temperatures
the effective mass starts to depend on $T$,
$M^*\propto T^{-1/2}$. This $T^{-1/2}$ dependence of the effective
mass at elevated  temperatures leads to the non-Fermi liquid behavior
of the resistivity. The application of a magnetic field $B$ restores
the common $T^2$ behavior of the resistivity.
A re-entrance into the highly correlated
regime is observed if the magnetic field $(B-B{c0})$ decreases to zero,
while the effective mass $M^*$ diverges as
$M^*\propto(B-B_{c0})^{-2/3}$.
At finite magnetic fields, the regime is restored at some temperature
$T^*\propto(B-B_{c0})^{4/3}$.
We have demonstrated
that this $B-T$ phase diagram has a strong impact on the
magnetoresistance (MR) of the highly correlated electron liquid. The
MR as a function of the temperature exhibits a transition from the
negative values of MR at $T\to 0$ to
the positive values at $T\propto (B-B_{c0})^{4/3}$.
Such a behavior was observed in the heavy fermion metal CeCoIn$_5$.

We conclude that
FCQPT can be viewed as a universal cause of the non-Fermi liquid
behavior observed in different metals and liquids.

\section{Acknowledgments}

This work was supported in part by the Russian Foundation
for Basic Research, project No 01-02-17189.

\end{document}